\documentclass[traditabstract]{aa}  

\usepackage{graphicx}
\usepackage{natbib}
\usepackage{lscape}
\usepackage{supertabular}
\usepackage{xcolor}

\newcommand{\vsini}{$\varv \sin{i}$}
\newcommand{\hei}[0]{\ion{He}{i}}

\usepackage{txfonts}

\begin{document}

\title{The CARMENES search for exoplanets around M dwarfs}
\subtitle{The \ion{He}{i} triplet at 10830 \AA\, across the M~dwarf sequence\thanks{Full Table 2 is only available in electronic form
at the CDS via anonymous ftp to cdsarc.u-strasbg.fr (130.79.128.5)
or via http://cdsweb.u-strasbg.fr/cgi-bin/qcat?J/A+A/}}

\author{B. Fuhrmeister\inst{1}, S. Czesla\inst{1}, L. Hildebrandt\inst{1}, E. Nagel\inst{1}, J. H. M. M. Schmitt\inst{1}
  \and  D.~Hintz\inst{1}
  \and  E.~N.~Johnson\inst{2}
  \and  J.~Sanz-Forcada\inst{3}
  \and  P.~Sch\"ofer\inst{2}
  \and  S.~V.~Jeffers\inst{2}
  \and  J.~A.~Caballero\inst{3}
  \and M.~Zechmeister\inst{2}
  \and   A.~Reiners\inst{2}
  \and   I.~Ribas\inst{4,5}
  \and P.~J.~Amado\inst{6}
  \and  A.~Quirrenbach\inst{7}
  \and  F.~F.~Bauer\inst{6}
  \and  V.~J.~S.~B\'ejar\inst{8,9}
  \and  M.~Cort\'es-Contreras\inst{3}
  \and  E.~D\'iez-Alonso\inst{10,11}
  \and  S.~Dreizler\inst{2}
  \and  D.~Galad\'{\i}-Enr\'{\i}quez\inst{12}
  \and  E.~W.~Guenther\inst{13}
  \and  A. Kaminski\inst{7}
  \and  M.~K\"urster\inst{14}
  \and  M.~Lafarga\inst{4,5}
  \and  D.~Montes\inst{10}}

\institute{Hamburger Sternwarte, Universit\"at Hamburg, Gojenbergsweg 112, D-21029 Hamburg, Germany\\
  \email{bfuhrmeister@hs.uni-hamburg.de}
        \and
        Institut f\"ur Astrophysik, Friedrich-Hund-Platz 1, D-37077 G\"ottingen, Germany 
        \and
        Centro de Astrobiolog\'{\i}a (CSIC-INTA), ESAC, Camino Bajo del Castillo s/n, E-28692 Villanueva de la Ca\~nada, Madrid, Spain 
        \and
        Institut de Ci\`encies de l'Espai (ICE, CSIC), Campus UAB, c/ de Can Magrans s/n, E-08193 Bellaterra, Barcelona, Spain
        \and
        Institut d'Estudis Espacials de Catalunya (IEEC), E-08034 Barcelona, Spain
           \and 
        Instituto de Astrof\'isica de Andaluc\'ia (CSIC), Glorieta de la Astronom\'ia s/n, E-18008 Granada, Spain 
        \and 
        Landessternwarte, Zentrum f\"ur Astronomie der Universit\"at Heidelberg, K\"onigstuhl 12, D-69117 Heidelberg, Germany 
        \and
        Instituto de Astrof\'{\i}sica de Canarias, c/ V\'{\i}a L\'actea s/n, E-38205 La Laguna, Tenerife, Spain
        \and
        Departamento de Astrof\'{\i}sica, Universidad de La Laguna, E-38206 Tenerife, Spain 
        \and
        Departamento de F\'isica de la Tierra y Astrof\'isica \& IPARCOS-UCM (Instituto de F\'sica de Part\'culas y del Cosmos de la UCM), Facultad de Ciencias F\'isicas, Universidad Complutense de Madrid, E-28040 Madrid, Spain 
        \and
        Departamento de Explotaci\'on y Prospecc\'on de Minas, Escuela de Minas, Energ\'{\i}a y Materiales, Universidad de Oviedo, E-33003 Oviedo, Asturias, Spain
        \and
         Observatorio de Calar Alto, Sierra de los Filabres, E-04550 G\'ergal, Spain 
        \and
        Th\"uringer Landessternwarte Tautenburg, Sternwarte 5, D-07778 Tautenburg, Germany 
        \and
        Max-Planck-Institut f\"ur Astronomie, K\"onigstuhl 17, D-69117 Heidelberg, Germany 
       }
        
\date{Received dd/10/2018; accepted dd/mm/2018}

\abstract
    {The \hei\,  infrared (IR) triplet at 10830 \AA\, is an important activity indicator for the Sun
      and in solar-type stars, however, it has rarely been studied in relation to M~dwarfs to date.
      In this study, we use the time-averaged spectra 
      of 319 single stars with spectral types ranging from M0.0\,V to M9.0\,V obtained with the CARMENES high resolution optical and near-infrared spectrograph at Calar Alto
  to study the properties of the \hei\, IR triplet lines.
  In quiescence, we find the triplet in absorption with a decrease of the
  measured pseudo equivalent width (pEW) towards later sub-types.
  For stars later than M5.0\,V, the \hei\, triplet becomes undetectable in our study.
  This dependence on effective temperature may be related to a change in
  chromospheric conditions along the M~dwarf sequence.
  When an emission in the triplet is observed, we attribute it to flaring.
  The absence of emission during quiescence is consistent with 
  line formation by photo-ionisation and recombination, while flare emission
  may be caused by collisions within dense material.
  The \hei\ triplet tends to increase in depth according to increasing activity levels, ultimately becoming filled in; however, we do not
  find a correlation between the pEW(He IR) and X-ray properties.
  This behaviour may  be attributed to the absence of very inactive stars ($L_{\rm X}/L_{\rm bol}$ < -5.5)
  in our sample or to the complex behaviour with regard to increasing depth and filling in. 
    }

\titlerunning{Activity in the \ion{He}{i} IR line}
\authorrunning{B. Fuhrmeister et~al.}
\maketitle


\section{Introduction}

Studies of stellar chromospheres typically rely on observations of chromospherically
sensitive lines, such as the \ion{Ca}{ii} H \& K lines or the H$\alpha$ line. 
In the infrared (IR) regime,
the \ion{He}{i} IR triplet lines at 10830~\AA\ are a valuable indicator of chromospheric
activity. 
The \ion{He}{i} triplet is
formed by transitions from the meta-stable $2^{3}\mathrm{S}$ level
to the $2^{3}\mathrm{P}$ level, the central wavelengths of the transitions being located at
10\,832.057, 10\,833.217, and 10\,833.306~\AA. We note that these and all other wavelengths 
are given for vacuum conditions since the data used here are measured in vacuum.
The latter two components dominate the triplet and remain unresolved in the majority of studies, which is
why we refer to the blend of these components as the abbreviation: \hei\ IR line.

The \ion{He}{i} triplet lines have frequently been used in solar studies where
the lines are strongly modulated by the solar activity cycle \citep{Livingston}.
They have also extensively been used for chromospheric studies, such
as the determination of the magnetic field in solar prominences by spectropolarimetric
observations \citep{Orozco}.  In the stellar context,
the \hei\, IR line has widely been used to study young stars, 
specifically T Tauri stars, where the triplet is found to be a broad emission component with 
narrow or broad absorption features. While the emission is thought to originate in the post-shock
regions \citep{Dupree2014}, the absorption features serve as a probe of winds and
in-falling material around accreting T~Tauri stars \citep{Dupree2005, Edwards2006}, as well as
more massive Herbig Ae/Be stars \citep{Cauley}.  Recently, the
\hei\, IR triplet has also been found to be a highly informative tracer of the outer
atmospheres of exoplanets \citep[e.g.,][]{Spake2018, Nortmann2018,
Salz2018, Allart2018, Mansfield2018, Alonso-Floriano2019}.

Early systematic measurements of the \ion{He}{i} triplet lines as a chromospheric indicator in 
main sequence stars 
date back to \citet{Zirin1968}, who  frequently found the triplet lines in
absorption for G and K dwarfs. \citet{Zirin1982} extended this study by
also identifying the line  in some early M~dwarf stars. Since then, the line has
been extensively used as a chromospheric diagnostic  for solar-type stars. For example, \citet{Andretta2017} used the line ratio between \ion{He}{i} IR and
\ion{He}{i} D$_{3}$ (at 5877.24 \AA) to infer filling factors for active regions of solar-like
stars through their comparison to solar models. 

The formation scenario of the \ion{He}{i} IR triplet lines in the atmospheres of main 
sequence and giant stars has been under debate for more than 30~years.
The meta-stable ground level of the triplet transitions can be populated by 
photo-ionisation and recombination (PR)
processes, requiring the presence of photons with wavelengths of 504~\AA\ or 
below to ionise helium. The meta-stable
level is then populated by recombining and downward-cascading electrons.
Alternatively, in rather dense chromospheric layers, collisional processes can also populate the meta-stable level. 
To distinguish between these two scenarios, \citet{Zarro1986} studied a sample of 
about 70 dwarfs with spectral types between K3 and F0 and found
a correlation between the ratio of X-ray and bolometric surface fluxes,
$\log(f_{\rm X}/f_{\rm bol})$, and the equivalent widths (EW) of the  \hei\, line for dwarfs later 
than (and including) F7, leading \citet{Zarro1986} to the conclusion that the PR mechanism
makes predictions that are, at least, qualitatively consistent with their observations.

\citet{Takeda2011} observe the \hei\,IR line in a sample of 33 metal-poor late-type stars, of 
which 24 were dwarfs of type G or hotter, and measure the EWs of the line, EW(He~IR), using
Gaussian fits. In their dwarf sample, the authors find an EW of $30$~m\AA\ for 
all low-metallicity dwarfs with \mbox{[Fe/H]$<-1$}, while stars with higher metallicity 
demonstrated a larger scatter in their EWs. In combining this data with the findings of \citet{Zarro1986}, 
\citet{Takeda2011} find a satisfactory correlation between EW(He IR) and $\log(f_{\rm X}/f_{\rm bol})$.
\citet{Smith2016} studies the work of \citet{Zarro1986} which is extended by a number of stars.
Like \citet{Zarro1986}, \citet{Smith2016} finds a satisfactory correlation between \hei\,IR line absorption with 
fractional X-ray luminosity, $L_{\rm X}/L_{\rm bol}$,
along\ with \ion{Ca}{ii} H\&K emission for their later-type sub-sample.
However, neither of these samples have included any M~dwarfs.

\citet{Dupree2018} study the \hei\, line in a sample of 11 dwarfs 
with spectral types between G0.5\,V and M5\,V, observed with
the PHOENIX spectrograph mounted at the 4~m Mayall telescope, and 
find a decrease in EW(He IR) with
spectral type for stars later than about K. Further, \citet{Dupree2018} argued that the  
EW(He IR) is related to the surface X-ray flux for F- through
early M-type stars. The latest M~dwarf stars in their sample demonstrated the largest surface X-ray fluxes,
but only with weak \hei\,IR lines.

While all these correlations between EW(He IR) and X-ray flux can be interpreted
in favour of the PR mechanism, they may, alternatively, be caused by a general activity trend.
In a theoretical study, \citet{Andretta1997} showed that there can be a complex interplay between
the PR mechanism and collisional excitation.
\citet{Sanz-Forcada2008} observe active dwarfs, sub-giants, and giants, finding
a correlation between EW(He IR) and extreme ultraviolet
(EUV) and X-ray fluxes for giants and dwarfs of low activity, but not for the most active dwarfs. 
These observations indicate that the PR process dominates in giants, while the
collisional effects become more relevant in the denser chromospheres of
active dwarf and sub-giant stars.  

A more detailed picture is obtainable from solar data, where
spatially-resolved flare observations by \citet{Kobanov}
can be interpreted in terms of the PR process. In particular, the \hei\,IR line
was strengthened in terms of absorption, while other chromospheric and coronal lines
brightened. Nevertheless, the observations showed a time lag with
inconclusive interpretation
between the emission maximum in the EUV and the absorption maximum in the
\hei\,IR line for a second flare maximum.
Although many findings favour the PR mechanism,
the issue of the appropriate line formation scenario has not been settled.

In M dwarfs, the behaviour of the \hei\,IR line has remained largely unexplored to date.
During flares, the \hei\,IR line is observed going into emission along
with other helium, Balmer, Paschen, and Bracket lines by, for example, \citet{Schmidt2012}, yet
detections in absorption remain quite rare since only a few authors include M dwarfs
in their studies \citep{Zirin1982, Dupree2018} despite the M dwarf sequence exhibiting
some major changes in other chromospheric lines. For example, the H$\alpha$ line emission
peaks around M5\,V and then declines to an even later spectral type
due to the increasing neutrality of these ultra-cool atmospheres \citep{Mohanty}.  The  question arises, thus, about whether the \hei\,IR line exhibits a similar behaviour.

Here we use a sample of over 300 M~dwarfs observed and monitored in the context
of the Calar Alto high-Resolution search for M dwarfs with Exo-earths with Near-infrared and optical 
Echelle Spectrographs (CARMENES) survey to perform a comprehensive study of the
\hei\,IR line throughout this spectral type regime. Also using CARMENES data,
\citet{Patrick} perform a more general activity study of M dwarfs, including
the \hei\, line, but concentrating on other lines and using an entirely different measuring technique. We provide a comparison with this study in Sect. \ref{previous}.

Our paper is structured as follows: in Sect.~2, we give an overview of the data used; in Sect.~3 we describe our pEW measurement method using fits with Voigt profiles and we 
present our results; a discussion in Sect.~4; and our conclusions in
Sect.~5.

\section{Observations, sample selection, data reduction, and telluric correction}

All spectra used for the present analysis were obtained with 
CARMENES \citep{Quirrenbach2018}, mounted at the 3.5\,m Calar Alto telescope.
CARMENES is a
two-channel, fibre-fed spectrograph covering the wavelength range from 0.52\,$\mu$m to 0.96\,$\mu$m 
in the visual channel (VIS) and from 0.96\,$\mu$m to 1.71\,$\mu$m in the near infrared
channel (NIR) with spectral resolutions of 94\,600 and 80\,400 in the VIS and NIR, respectively. 
The CARMENES consortium is monitoring more than 300 M~dwarfs in a comprehensive search for low-mass exoplanets in
their habitable zones \citep{AF15a, Reiners2017}.  To date, CARMENES has obtained more than 
14\,000 high-resolution visible and near-infrared spectra in the context of this ongoing survey.
All spectra are consistently reduced with a dedicated CARMENES reduction pipeline
\citep{pipeline}.  

Our study of chromospheric activity visible in the \hei\,IR line
is based on the spectra of the 335 stars listed in the CARMENES data archive as of 1 Feb 2019, for which both VIS
and NIR spectra are available.  After excluding 16 known spectroscopic binaries \citep{carmencita, Baroch2018}, we are left with a sample of 319 M-type stars. 
This collection of 319 stars is referred to as the whole sample or all 
sample stars in the following.

\begin{figure*}
\begin{center}
\includegraphics[width=0.5\textwidth, clip]{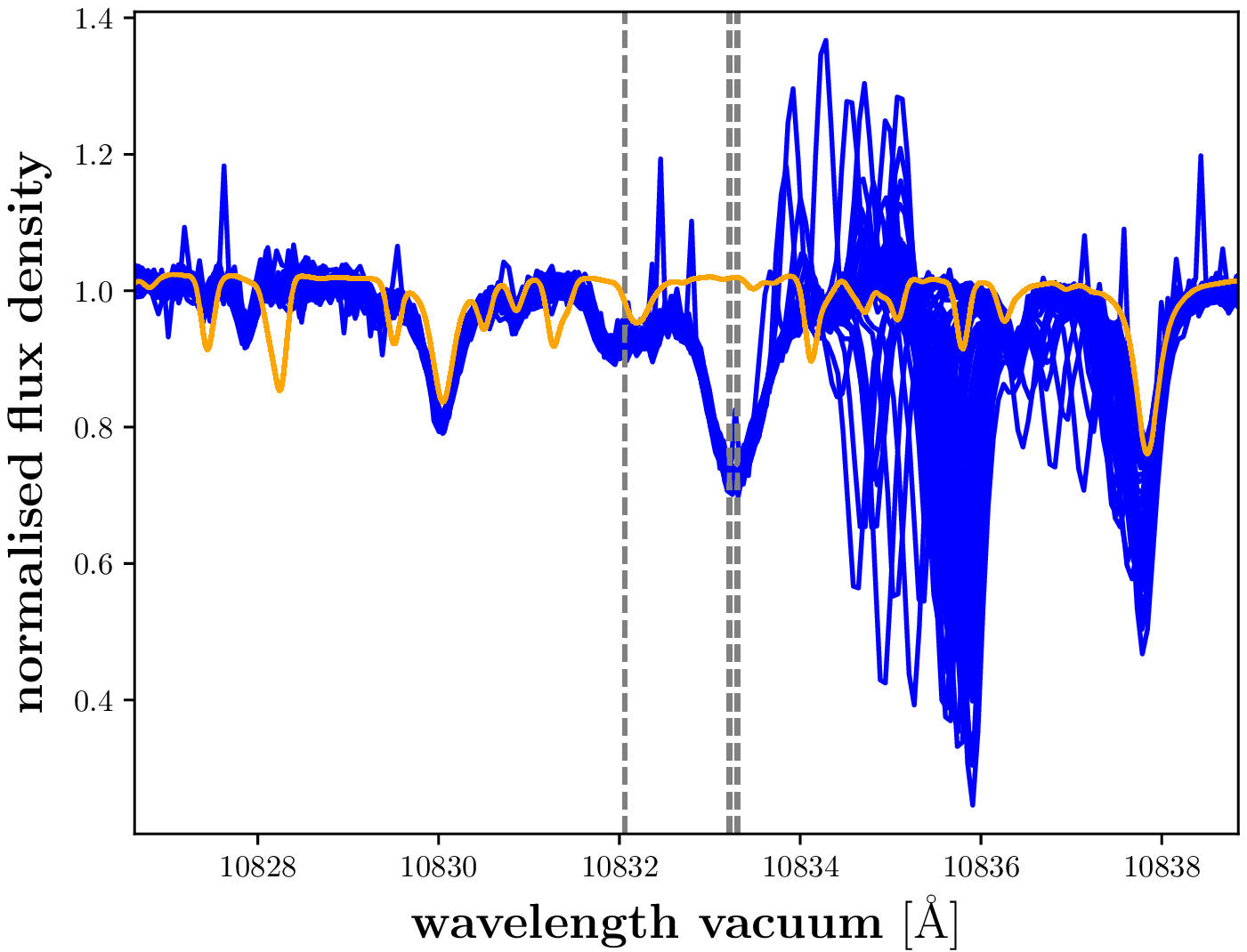}
\includegraphics[width=0.5\textwidth, clip]{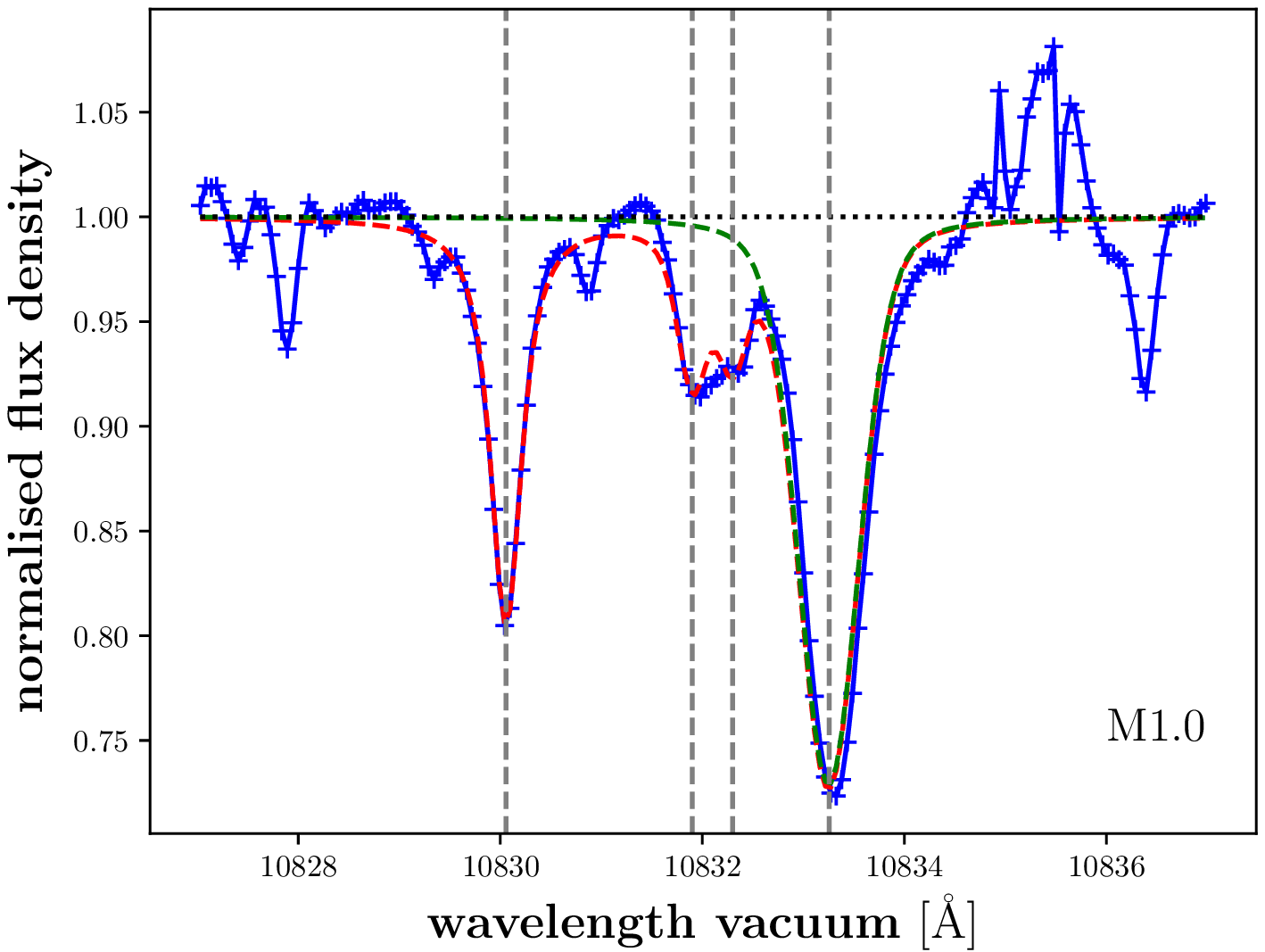}
\caption{\label{spectrum} \emph{Left}: All spectra of J00051+457/GJ~2 used in our analysis in blue.
  The yellow spectrum is a PHOENIX photospheric model spectrum for comparison. \emph{Right}: Co-added and telluric corrected spectrum of J00051+457/GJ~2 in blue with best fit Voigt model in red. The green dashed
  line represents the \hei\ line component of the fit. At $\sim$10\,835~\AA,\ some artefacts of
  telluric airglow lines can be seen (shown here in the stellar rest frame).
  In the left panel, we show the positions
  of the three components of the \hei\, IR line as dashed grey lines.
  In the right panel, the grey dashed vertical lines mark the positions of  lines
  considered in the fit (see text for details). }
\end{center}
\end{figure*}

As an example of the CARMENES spectra of the \hei\,IR region, we show 
in Fig.~\ref{spectrum} (left panel) the 52 available NIR spectra of the M1.0\,V star J00051+457/GJ~2 
before telluric correction and normalised around the \hei\ triplet lines.
The CARMENES spectral time series are obtained at a temporal cadence optimised for
planet detection. Therefore the typical time differences between consecutive observations
vary from a day to a few weeks, and only rarely multiple exposures on a given night are available.
In the specific case of J00051+457/GJ~2 (shown in Fig.~\ref{spectrum}), 
all spectra were taken between July 2016 and December 2017, but for
most of our sample stars, there are also more recent spectra. 
The CARMENES spectral time series are suitable for
searches of periodic variations in the chromospheric lines, which we demonstrate
in \cite{rotation}. Yet the amplitude of rotational
modulation is typically quite low in H$\alpha$ and  \ion{Ca}{ii} IRT lines, favouring photometric
studies to determine rotational periods.

In this respect, the star J00051+457/GJ~2 is a rather typical example of an early M~dwarf.
As can be seen in Fig.~\ref{spectrum}, it neither demonstrates strong rotational modulation nor
considerable flaring activity in the \hei\,IR line, a result that also holds for other chromospheric
lines such as H$\alpha$. Although J00051+457/GJ~2 shows H$\alpha$ in absorption and would therefore 
formally be classified as inactive, it is not a particularly inactive representative of our sample,
which is also indicated by its emission cores in the \ion{Ca}{ii} H\&K lines \citep{rauscher2006}.

As also shown in Fig.~\ref{spectrum}, many telluric absorption and emission lines
are found in the region around the \hei\,IR triplet, with their strengths capable of considerably exceeding 
the sought-after stellar \hei\,IR line.   Furthermore, since the CARMENES spectra are corrected
for barycentric velocity shifts, the telluric lines usually pass through the spectra of a given
target.   The most relevant water absorption lines in the \hei\,IR triplet region \citep[see][]{Rothman2012, HITRAN2016}
are located at 10\,829.69, 10\,833.32, 10\,834.59, 10\,835.07, and 10836.94~\AA.
To correct for telluric absorption
lines, we applied the method described by \citet{Evangelos}
using \texttt{molecfit} \citep[][]{molecfit, molecfit2}\footnote{\texttt{https:\newline//www.eso.org/sci/software/pipelines/skytools/molecfit}}.
This technique accounts for all water absorption lines in the region,  specifically including the ones quoted.   

As for airglow lines, upon comparison with the list compiled by \citet{Oliva2015}, we identify
airglow lines at wavelengths of 10\,832.103, 10\,832.412, 10\,832.271, 110834.241, and
10834.338~\AA\, originating in OH; the latter two are not resolved in the CARMENES spectra.
To account for the airglow lines, we took 
particular advantage of the time-variable barycentric shift of the telluric lines 
with respect to the stellar spectrum.
Based on the known wavelengths of the airglow lines, a mask with airglow-affected regions was constructed.
Using the Spectrum Radial Velocity AnaLyser (SERVAL) software \citep{serval}, we then
co-added the available spectra of each individual star in the barycentric rest frame,
substantially down-weighting the masked regions. As a result of this procedure, the impact of the airglow lines
is significantly reduced.

%

The application of the spectral co-addition of the telluric
corrected spectra by the SERVAL software produces an averaged spectrum for each star. 
In Fig.~\ref{spectrum} (right panel), we show the 
resulting averaged spectrum for J00051+457/GJ~2, 
thus demonstrating our ability to remove the effects of tellurics in our spectra.
The averaged spectra are essentially free of telluric absorption lines, althought they may sometimes display artefacts 
from airglow lines. The quality of the resulting correction for the airglow lines depends, in fact, 
on the number of available spectra and their distribution in barycentric velocity;
for the typical sampling of the CARMENES spectra \citep{Garcia-Piquer}, we find that ten spectra are
usually sufficient to obtain a satisfactory average spectrum for the star.
In the VIS channel of CARMENES, telluric contamination is far less important, 
especially near the chromospherically active lines relevant to our study, namely, the H$\alpha$ line,
the \ion{Ca}{ii} infrared-triplet (IRT) lines, and the \ion{He}{i} D$_{3}$ line.  In this case,
we find that a bin-wise median of the VIS channel spectra yields satisfactory results.

In this paper, we focus on the average spectral properties of our sample. 
Similar to the case of J00051+457/GJ~2, we verified by visual inspection that, in particular, the 
earliest M dwarfs in our sample don't  show pronounced variability in the \hei\, IR nor in the 
H$\alpha$ line.   While we, therefore, consider our time-averaged analysis satisfactory in most cases, 
temporally-resolved analyses are required in individual cases. In the following, we point out cases where
variability may have a significant effect on our analysis but we will save a detailed analysis of all time-dependent effects for further discussion in the future.

\section{Equivalent width measurements of chromospheric lines}

The CARMENES VIS and NIR  spectra, averaged as described above and corrected for telluric absorption
and emission lines for all 319 M dwarfs form the basis of our study.
The spectra of M~dwarfs do not show an identifiable continuum 
because of the ubiquitously present molecular absorption lines; therefore, we employ the method of
pseudo equivalent widths (pEWs).  The individual steps of our procedure and our results
are described in the following sections.

\subsection{Photospheric models of the \hei\,IR triplet line region}

The strong photospheric background makes any pEW estimation in the region of the \hei\,IR triplet
lines challenging.  To illustrate the difficulties, we may consider once again the spectrum of the M1~V star
J00051+457/GJ~2
(Fig.~\ref{spectrum}, left panel), where a photospheric model
spectrum, computed with PHOENIX \citep{Hauschildt1999} with an effective temperature of
$T_\mathrm{{eff}} = 3700$ K, $\log{g} = 5.0$,
and solar chemical composition \citep[from the grid by][]{Husser} is overplotted.
This most recent PHOENIX spectral library uses local thermal equilibrium (LTE)
calculations of atomic and
molecular lines in a spherical, one-dimensional geometry.  As can be seen from Fig.~\ref{spectrum} (left panel),
this model reproduces the prominent atomic \ion{Si}{i} and \ion{Na}{i} lines
at 10\,830.057~\AA\, and 10\,837.814~\AA\, rather well, while
other absorption lines are not well reproduced. These lines are likely of 
molecular origin since, typically, the available molecular line data are far less
accurate than atomic line data.
Moreover, the observed spectrum of J00051+457/GJ~2 shows a pronounced absorption feature 
at about $10833$~\AA, which is also not reproduced by the model.  We identify this feature 
with the blend of the two reddest and strongest components of the \hei\,IR triplet lines, 
which do not originate in the photosphere, but rather in the chromosphere. Since PHOENIX
calculates only the photospheric emission, but not the chromospheric emission, the feature
isn't included in the models. This applies  to other chromospheric features as well. Just like
the \hei\, IR line, also the \hei\,D$_{3}$ and the H$\alpha$ lines are not present in the
PHOENIX photospheric models since they are chromospheric in origin. To a lesser extent, this
also holds\ for the \ion{Ca}{ii} IRT lines, which demonstrate a much stronger contribution from the
photosphere.

\subsection{Photospheric lines around 10830\,\AA\, across the M-type regime}
\label{sec:photlines}

\begin{figure}[htb]
\begin{center}
\includegraphics[width=0.5\textwidth, clip]{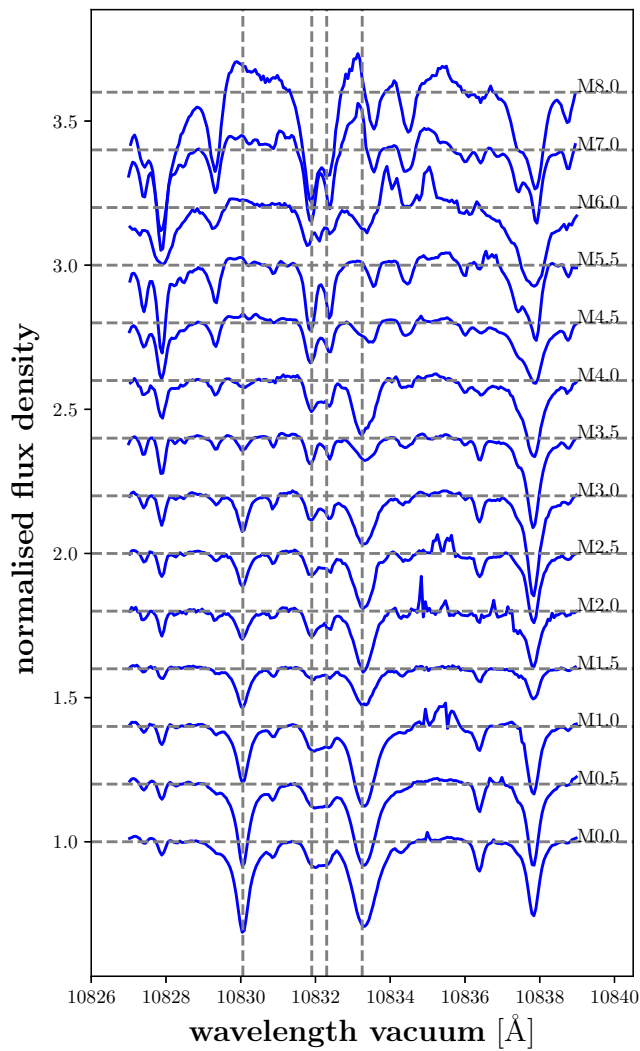}
\caption{\label{earlyandlateMdwarf} CARMENES spectra of spectral region around \hei\,IR line
  across M dwarf spectral sequence exhibiting considerable change from M0 to M8. Each
  normalised spectrum is offset by 0.2 in flux density for the purposes of clarity.
  The observed normalised spectrum is denoted in blue with its local pseudo-continuum/offset
  marked as a dashed grey horizontal line.
  Vertical grey dashed lines mark the position of the lines
  considered in the fit. The stars shown are
  M0.0\,V: J03463+262/HD~23453;
  M0.5\,V: J02222+478/BD+47~612;
  M1.0\,V: J00051+457/GJ~2;
  M1.5\,V: J02123+035/BD+02~348;
  M2.0\,V: J01013+613/GJ~47;
  M2.5\,V: J00389+306/Wolf~1056;
  M3.0\,V: J02015+637/G~244-047;
  M3.5\,V: J12479+097/Wolf~437;
  M4.0\,V: J01339-176/LP~768-113;
  M4.5\,V: J01125-169/YZ~Cet;
  M5.5\,V: J00067-075/GJ~1002;
  M6.0\,V: J14321+081/LP~560-035; 
  M7.0\,V: J02530+168/Teegarden's star; and
  M8.0\,V: J19169+051S/vB10.
  }
\end{center}
\end{figure}

In the spectral range from  early to late M dwarfs, the photospheric spectrum
within which the chromospheric \hei\,IR line is embedded changes considerably,
as demonstrated in Fig.~\ref{earlyandlateMdwarf}.
For the early M~dwarfs, the \ion{Si}{i} line at 10830.057~\AA\, 
dominates the photospheric spectrum,
yet this line weakens towards later sub-types and essentially vanishes at a spectral type
around M4. The \ion{Na}{i} line at 10\,837.814 \AA\, starts to deform around that spectral type
due to its blending with a line that is, most probably, molecular. 

In the wavelength range between these two atomic lines, more
spectral features can be seen, which we believe are likely to be  molecular in origin, 
except for the \hei\, line at 10\,833.3\,\AA. 
Molecular lines gain in strength and number when approaching later spectral types, in
contrast to many atomic lines. Moreover, none of these lines is appropriately represented in the PHOENIX
model spectrum shown in Fig.~\ref{spectrum}. This is typically the case for molecular lines from
lesser-known transitions, while atomic data are often known to a much greater extent, as 
demonstrated by the well-reproduced two atomic lines in the PHOENIX model spectrum.
We cannot count out the possibility that unidentified atomic lines also contribute to the spectrum, but we emphasise
that the ansatz adopted in our study of the
\hei\,IR line does not rely on accurate knowledge of the
origin of these lines (see Sect.~\ref{pewestimation}).

Between spectral types M3.5\,V and M4.5\,V, an unidentified line emerges
on the red side of the \hei\, line, which becomes stronger towards even later spectral
types and is, thus, likely to be of molecular origin. This line can lead to severe blending
with the \hei\,IR line, producing triangular shapes for the line, 
which may complicate the detection and analysis of the \hei\, line.
Fortunately, such cases of extreme blending  are quite rare. Through visual inspection, we
find only seven examples of lines with triangular shapes in our sample, all 
for stars with spectral types later than M4.0. The molecular line emergence can be seen in more detail in
Fig.~\ref{spectralseries2} from the middle left panel to bottom right panel. The bottom left
panel exhibits an example of a triangular line shape for the M6.0\,V star J14321+081/LP~560-035.
In this case, the fit most likely overestimates
the \hei\,IR line strength because of the unresolved blend.

More importantly, the dominance of unidentified lines 
leads to considerable difficulties in the estimation of the level of the pseudo-continuum for
our M~dwarfs later than about M5.5\,V.
To give a better overview of the fit quality across the spectral sequence, we show fits
of the \hei\,IR line for the stars used in Fig.~\ref{earlyandlateMdwarf}
in Figs.~\ref{spectralseries1} and \ref{spectralseries2} in Appendix~\ref{appendixa}. 


\subsection{pEW estimation for the \hei\,IR triplet lines}\label{pewestimation}

Since the model spectra do not provide any appropriate reference, we applied the following procedure
to measure the pEW of the \hei\,IR triplet lines for the averaged and telluric corrected
spectra.
First, we fit the spectra in the wavelength range between 10\,829 to 10\,835\,\AA\, with an 
empirical model consisting of four Voigt profiles to reproduce the pertinent spectral lines in this range;
their starting values and free parameters are summarised in Table~\ref{Voigtpar}.
The component Voigt~1 accounts for the \ion{Si}{i} line at 10\,830.057~\AA, and the component Voigt~4
accounts for the unresolved lines of the \hei\,IR triplet at
10\,833.217, and 10\,833.306~\AA, which are treated as a single line centred at 10\,833.25 \AA.
The components Voigt~2 and Voigt~3  account for the unidentified spectral line components
discussed in Sect.~\ref{sec:photlines} and they are centred at 10\,831.9~\AA\ and 10\,832.3 \AA.
The component Voigt~3 is centred
near the weakest component of the \hei\,IR triplet
at 10\,832.057~\AA. However, this triplet component is essentially always blended with
the stronger, unidentified feature.

We fit the strengths of all lines and the shape parameters of the
component representing the strong \ion{Si}{i} line and the two strongest components
of the \hei\,IR triplet.  The shape parameters of the two components representing the
two unidentified features remain fixed because we found this setting most effective in stabilising the fit 
results considerably. The wavelengths of the components are not varied in our fits. 
Finally, the pEW of the \hei\,IR line is obtained by integrating the 
component Voigt~4. Therefore, our pEW values do not refer to the entire triplet. Negative values indicate emission lines, while positive values correspond
to absorption lines.

As an example of our procedure, we refer again to the average spectrum of
J00051+457/GJ~2 in Fig.~\ref{spectrum} (right panel) along with our best-fit model described above. 
While, admittedly, not all spectral lines in the range can be reproduced reasonably well by our model,
the silicon and helium lines are  modelled particularly well and, therefore, 
we consider the chosen approximation sufficient to
obtain an accurate estimate of the pEW for the \hei\,IR line. 

To obtain an error estimate for the
pEW, we vary the strength of the best-fit
component Voigt~4 and compute the resulting (larger) $\chi^{2}$ values.
Then we fit the resulting $\chi^{2}$ curve with a second-order polynomial.
Finally we determine
for which pEW the $\chi^{2}$ values increases by unity
compared to the best-fit value, which
yields an estimate of the uncertainty.

\begin{table}
\caption{\label{Voigtpar} Parameters for Voigt fit$^{a}$. }
\footnotesize
\begin{tabular}[h!]{lcccc}
\hline
\hline
\noalign{\smallskip}

Parameter               & Voigt 1 & Voigt 2 & Voigt 3 & Voigt 4 \\
                        & \ion{Si}{i}& unidentif. & unidentif.& \ion{He}{i}\\
                        &            &              &             & 2red comp.\\
\hline
\noalign{\smallskip}
Amplitude               &  -1.0 (free)   & -1.0 (free)    & -1.0 (free)    &  -1.0 (free)   \\
Gaussian\\
width $\sigma$          & 0.1 (free)    & 0.1      & 0.1   & 0.1  (free)     \\
Lorentz scale \\
parameter $\gamma$      & 0.1 (free) &0.1    &  0.1  & 0.1 (free)        \\
Central\\
wavel. [\AA]            &10\,830.057& 10\,831.9 & 10\,832.3& 10\,833.25 \\

\noalign{\smallskip}
\hline

\end{tabular}
$^{a}$ The wavelength span of the whole fit is 10\,829 to 10\,835 \AA. The continuum
is fit by a linear function.
\normalsize
\end{table}

\subsection{Validity of the pEW estimates for the \hei\,IR lines}

Clearly, the fitting procedure described above provides a formal value for the \hei\,IR line strength.
However, not all of the derived values are scientifically meaningful. Moreover, 
the question arises regarding which threshold in pEW(He IR) should be used to decide whether
the line is present.  Therefore, we visually inspected  the averaged spectra  for the presence of
the \hei\ line and 
then tried to find criteria to reproduce these findings more objectively.
To that end, we first  excluded values based on bad fits with a reduced $\chi^{2}$-value larger than four, and
next we discounted insignificant measurements by excluding pEW measurements 
that were less than 1.5 times the estimated error. 
The application of these criteria left us with a list of stars with detected \hei\ lines that come
quite close to the list of those identified through visual inspection. This procedure should also minimise the number of
false detections of the \hei\ line at the expense of more non-detections,
where indeed \hei\ lines may be present. The problem of ambiguous spectra
cannot be solved totally anyways; also for visual inspection there is a
number of mainly late-type M dwarf spectra where presence of the \hei\ line 
is ambivalent.
In the following, we refer to measurements that fulfil both criteria ($\chi^{2}$ < 4 and pEW(He IR)>1.5~$\sigma$)
as
valid measurements, and the sub-sample of 181 stars 
for which we derived valid measurements is referred to, accordingly, as a valid sample.
Thus, our valid sample comprises 57\% of the entire sample of stars and contains
detections of the \hei\,IR line with
pEW values in the range between $-0.27$ and $0.75$~\AA.

As an additional quality check, we inspected all of our spectral fits visually.  For most of the early-type M dwarfs,
the fit worked quite well (cf., Fig.~\ref{spectrum}, right panel), whereas for the late spectral sub-types, 
the fits run into trouble with regard to the pseudo-continuum, and noise artefacts tended to make an appearance more frequently.
We also compared our results to the study by \citet{Dupree2018} whose sample contains four M dwarfs.
Two of their objects are also included in our sample, that is,  J13536+776/NLTT~35712  and 
J11054+435/GJ~412A,  for which we derive a pEW(He IR) of
0.050~$\pm$~0.050 (which is therefore not among our valid measurements)
and 0.112~$\pm$~0.048~\AA, respectively. \citet{Dupree2018}
reported pEW(He IR) values of 0.067~\AA\, and 0.139~\AA, respectively.
Therefore, the derived values and the values found by \citet{Dupree2018}
agree within their margins of error.
This comparison also suggests that for some stars we may have indeed
obtained the correct pEW(He IR)
measurements, but which were nonetheless rejected by our conservative selection criteria.

\subsection{Overview of the derived pEW(\hei\,IR) estimates}

We present all our valid pEW measurements of the \hei\,IR line, the
$\chi^{2}$-values of the Voigt fits, and their errors in Table~\ref{allmeasurements} (full table
available from CDS). Besides the CARMENES identification number Karmn \citep{Reiners2017}, we provide the common name, the spectral type of each star as taken
from the Carmencita database \citep{carmencita}, the effective temperature $T_{\rm eff}$
as taken from \citet{Schweitzer2018}, and the $L_{\rm X}/L_{\rm bol}$ values, if known (see Sect. \ref{X-ray}).

Despite having applied rather conservative \hei\,IR line detection criteria, 
our sample may still have contained some false positives.  In response, we checked for outliers.
First, there is only one pEW measurement larger than
0.30 \AA, belonging to J01352-072/Barta~161~12. This star demonstrates by far the largest projected
rotation velocity (\vsini) in our sample, leading to extreme line broadening \citep{Reiners2017}. The \hei\,IR lines are 
blended with the neighbouring lines, which leads to the high
pEW value. Thus, although the \hei\ line is present in this star, we excluded the star from 
our sample.

Subsequently, while the majority of the (mean) \hei\, lines in the remaining sample are in absorption,
there are ten stars showing the (mean) \hei\,IR line in emission and which
fulfil our selection criteria.
Looking at both the corresponding averaged spectra and  
all available individual spectra, we found that
in the two cases of J19255+096/LSPM~J1925+0938 and J21348+515/Wolf~926, the result can be attributed 
to noise or telluric
artefacts. Consequently, we excluded these stars from
our sample as well.

For the remaining eight stars with an average \hei\,IR emission (J02088+494/G~173-039,
J05084-210/2M~J05082729-2101444, J06574+740/2MASS~J06572616+7405265,
J10196+198/AD~Leo, J11476+002/LP~613-049~A, J12156+526/StKM~2-809, J22518+317/GT~Peg,
J23548+385/RX~J2354.8+3831), we find that 
the line is variable. Many of these stars belong to moving groups and are associated with
the young disc as listed in the Carmencita data \citep{carmencita, Vera2019}, and they are, therefore,
expected to show high levels of activity. In most cases, the \hei\,
line region is rather flat in the bona fide quiescent state and the line goes into emission 
in one or more spectra, which leads to a net emission line in the averaged spectrum.
Since the \hei\ excursions are associated with
an enhancement in the amplitude of the H$\alpha$ line, we attribute this to flaring.
Moreover, the \hei\ emission tends to be quite broad. Assuming thermal broadening of the
\hei\, line for temperatures of about 20\,000--25\,000~K, we arrive at a Gaussian
width of about 0.3--0.4~\AA\, which is consistent with the profiles of the
absorption lines. The widths of the emission lines tend to be much greater, which may be caused
by such turbulent broadening as is observed during flares \citep[e.g.,][]{Linsky1989, Fuhrmeister2018}.
In any case, the emission lines in our sample of averaged spectra 
are not persistent features.
Here, we  focus on the averaged spectra and will devote
a detailed discussion of the variability of the \hei\, line in a forthcoming paper.

In conclusion, cases where the \hei\ in our averaged spectra 
is in emission can be explained by individual spectra exhibiting emission
caused by flaring or, possibly, the long term variability of these stars. Therefore, we excluded
such cases with a negative pEW(He IR) from the valid sample within the discussion of the 
temperature dependence of the \hei\,IR line strength in Sect.~4.1. We do consider them, 
however, in our discussion on activity in Sect.~4.5.

\begin{table*}
\caption{\label{allmeasurements} pEW measurements of different chromospheric lines for sample stars with valid pEW(\hei\,IR) measurement (full table at CDS)}
\footnotesize
\begin{tabular}[h!]{llcccccccccc}
\hline
\hline
\noalign{\smallskip}

Karmn         & Name &Spec. &T$_{\rm eff}$$^{b}$& pEW(He IR)$^{c}$   & $\sigma$(pEW(He IR))  & $\chi^{2}$& pEW(H$\alpha$) & pEW(Ca IRT) & pEW(He D$_{3}$)$^{d}$ & $L_{\rm X}/L_{\rm bol}$  \\
               &     & type$^{a}$ &[K] &[\AA]           & [\AA]                    &               & [\AA] & [\AA] & [\AA] \\
\hline
\noalign{\smallskip}
J00051+457&GJ 2& M1.0&3675&0.247&0.05&0.85&0.352&0.201&-0.051& ...\\
J00183+440&GX And& M1.0&3615&0.118&0.049&0.925&0.324&0.261&-0.047&-4.672\\
J00286-066&GJ 1012& M4.0&3398&0.076&0.042&2.648&0.138&0.208&-0.043&...\\
J00389+306&Wolf 1056& M2.5&3537&0.154&0.05&1.33&0.278&0.24&-0.06&-4.771\\
J00570+450&G 172-030& M3.0&3426&0.117&0.048&2.305&0.139&0.217&-0.059&...\\
J01013+613&GJ 47& M2.0&3529&0.138&0.049&1.13&0.248&0.238&-0.058&-4.650\\
J01019+541&G 218-020& M5.0&2900&0.121&0.05&1.014&-5.015&0.038&-0.543&...\\
J01025+716&BD+70 68& M3.0&3488&0.19&0.05&1.018&0.27&0.23&-0.054&-5.499\\
J01026+623&BD+61 195& M1.5&3805&0.273&0.05&1.108&0.263&0.177&-0.058&-5.098\\
J01339-176&LP 768-113& M4.0&3349&0.118&0.046&2.743&-1.611&0.135&-0.174&-3.346\\

\noalign{\smallskip}
\hline

\end{tabular}\\
$^{a}$ Taken from Carmencita \citep{carmencita}.\\
$^{b}$ Taken from \citet{Vera}.\\
$^{c}$ Positive values of all given pEWs correspond to an absorption line, while
negative values correspond to an emission line.\\
$^{d}$ Normalisation effects give slightly negative values even for no line present.
We  use a threshold of pEW(He D$_{3}$) < $-0.08$~\AA\ to accept a measurement as an emission line.
\normalsize
\end{table*}

\subsection{PEW measurements of the \ion{He}{i}~D$_3$, H$\alpha$, and \ion{Ca}{ii} IRT lines}\label{optlines}

In addition to the \hei\,IR triplet lines, we also obtained pEWs of H$\alpha$,
the bluest line of the \ion{Ca}{ii} IRT,  and the \ion{He}{i}~D$_3$ line at 5877~\AA.
At these wavelengths, telluric contamination plays a minor role, which 
simplifies the derivation of pEWs considerably.
We obtained pEWs for these lines by integrating the median spectrum in specific
wavelength intervals. In Table~\ref{ew}, we list the central wavelengths and full widths
of the adopted line integration intervals along with the pseudo-continuum reference bands.
We followed \citet{Robertson} and \citet{GomesdaSilva} in adopting a width $w$ of 
1.6~\AA\, for the (full) width of the H$\alpha$ line band. While the line
is normally narrower when seen in absorption,  emission lines
of slowly to intermediately fast-rotating M~dwarfs typically exhibit this width. The
line can be much broader in the fastest rotators and
during flares \citep[e.g.,][]{Linsky1989, Hawley2003, Fuhrmeister2018}. 
While there are only seven stars with $v\sin i > 35$~km\,s$^{-1}$ in our sample, flaring
can be more of a problem for, at least, the most active stars where frequent or vigorous
flaring will not average out any further. In those cases, the pEW(H$\alpha$) may be underestimated. 

For the bluest \ion{Ca}{ii} IRT line 
we adopted 0.5~\AA\, for the width of the line band
because this line is narrower than the H$\alpha$ line even when seen in emission. For the
\ion{He}{i} D$_{3}$ line we adopted an intermediate width of $1$~\AA, which is typical for stars
where it is seen in emission. In our analysis, we found no case where the line is clearly seen in absorption.
Our pEW estimates for the H$\alpha$, the bluest \ion{Ca}{ii} IRT line,
and the \ion{He}{i} D$_{3}$ line are also listed in Table \ref{allmeasurements}
for the valid sample defined above.

The CARMENES VIS spectra cover even more \hei\, lines, which have been seen as emission lines at least during flares
in M dwarfs \citep{proxcen, CNLeoflare}. In particular, there is
a line at 7065~\AA\, which results from an ortho-helium transition like the \hei\ D$_{3}$ and IR lines.
Two further lines at 7281 and 6678~\AA\, are caused by para-helium transitions. We screened all averaged
spectra for these lines, but we could not identify any of them neither in emission nor absorption.
Since the relative intensity of all three lines is
lower than that of the \hei\, D$_{3}$ line at 5877 \AA, these lines may be hidden in the molecular
pseudo-continuum; an example is shown in~Fig.~\ref{he6678} in Appendix~\ref{otherhelium}.

\begin{table}
\caption{\label{ew} Parameters of  EW calculation (all wavelengths in vacuum). }
\footnotesize
\begin{tabular}[h!]{lcccc}
\hline
\hline
\noalign{\smallskip}

               & Centr.   & Full  & Reference band 1 & Reference band 2 \\
               & wavel.   & width& \\
& [\AA] & [\AA] & [\AA] & [\AA]\\
\hline
\noalign{\smallskip}
H$\alpha$& 6564.62 & 1.6 & 6537.43--6547.92 & 6577.88--6586.37 \\
\ion{Ca}{ii}  & 8500.33 & 0.5 & 8476.33--8486.33 & 8552.35--8554.35\\
\ion{He}{i} & 5877.24 & 1.0 & 5870.00--5874.00& 5910.0--5914.0\\
\noalign{\smallskip}
\hline

\end{tabular}
\normalsize
\end{table}

\section{Discussion}

\subsection{Dependence of the \hei\, IR line on effective temperature and spectral type}

\begin{figure}[tb]
\begin{center}
\includegraphics[width=0.5\textwidth, clip]{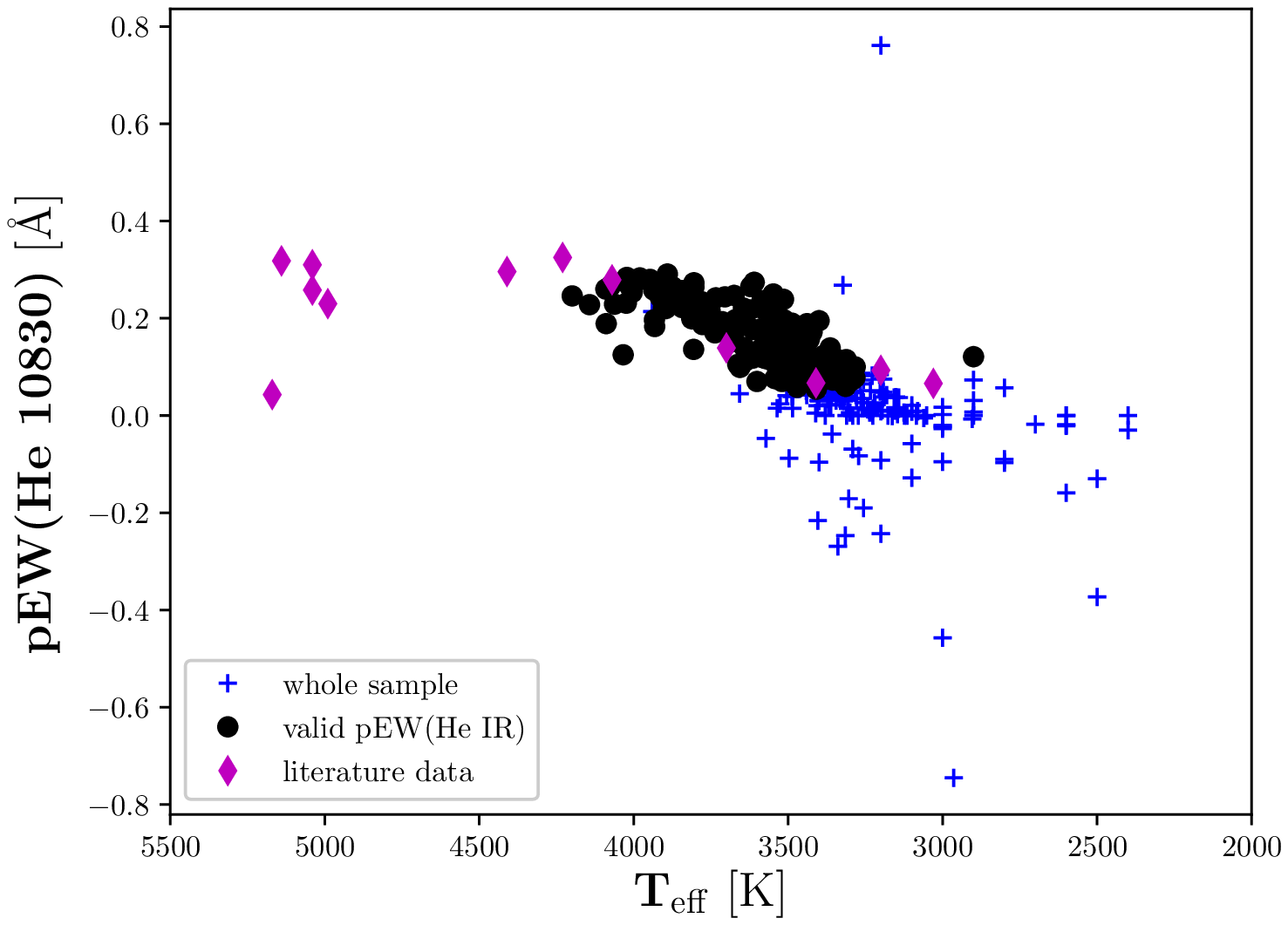}\\
\includegraphics[width=0.5\textwidth, clip]{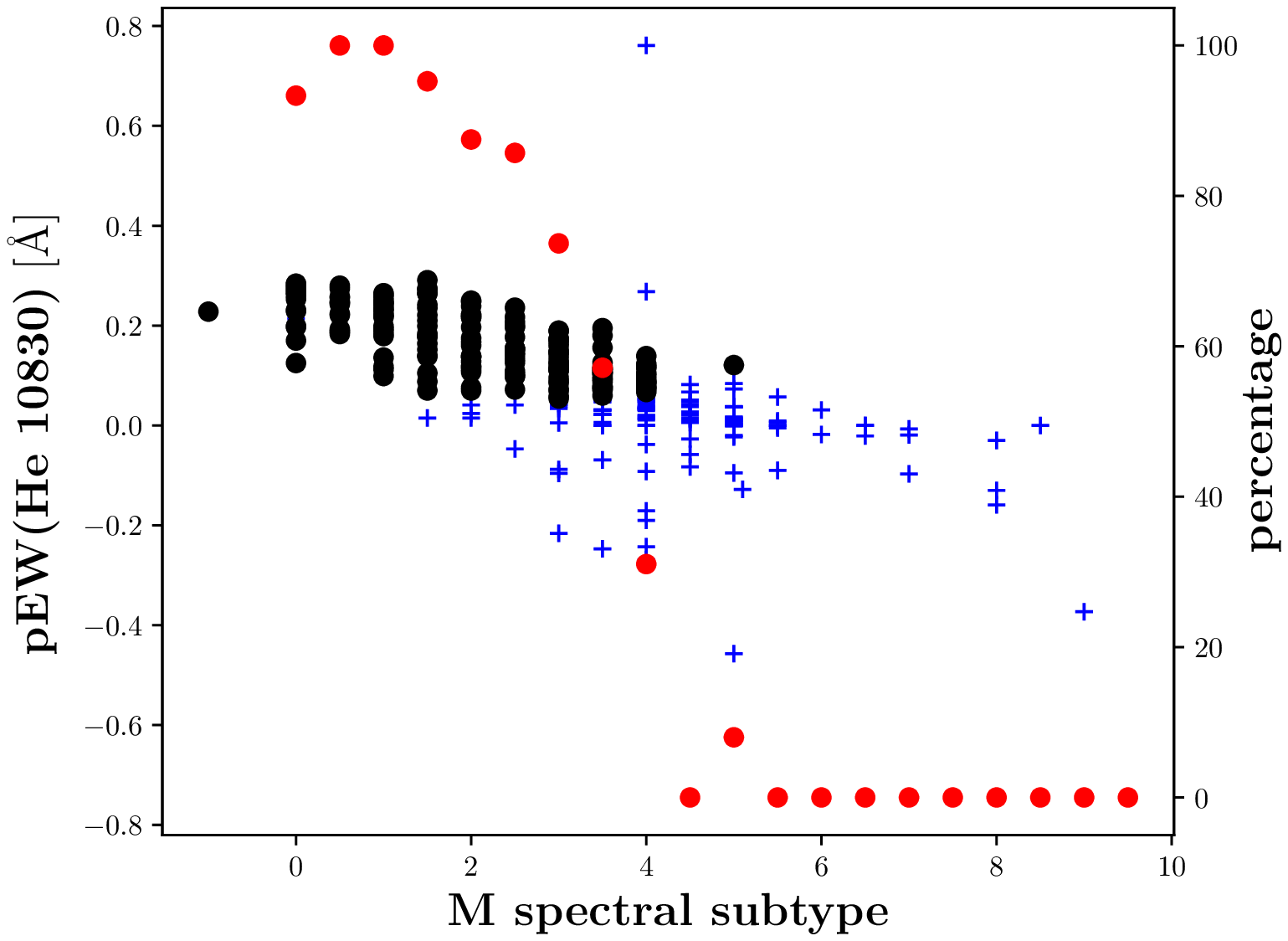}
\caption{\label{temperature} \emph{Top}: pEW(He IR) as function of effective
  temperature. 
  \emph{Bottom}: pEW(He IR) as function of spectral type.
  Blue crosses represent all sample stars (i.\,e. the non-valid pEWs), while black dots
  represent the as valid selected pEWs. In the top panel, magenta diamonds
  represent literature data from \citet{Sanz-Forcada2008}, \citet{Andretta2017}, and \citet{Dupree2018}. 
  In the bottom panel, red
  dots correspond to the right y-axis and give the percentage of valid measurements for the selection criteria,
  (see text).
  }
\end{center}
\end{figure}

We first consider the dependence of our pEW(He IR) measurements on the effective temperature $T_{\rm eff}$
and spectral type of the underlying star.   In the top panel of Fig.~\ref{temperature}, we plot our 
pEW(He IR) measurements as a function of effective temperature $T_{\rm eff}$, taken from 
\citet{Schweitzer2018}, who also use PHOENIX models for their stellar parameter determination.
As can be seen from Fig.~\ref{temperature}, 
the pEW declines towards lower effective temperatures and
no valid pEW measurements for the \hei\, line could be derived
for the stars with the lowest effective temperatures.  Besides, the range of our
pEW(He IR) measurements agrees well with corresponding values taken from the literature 
\citep{Dupree2018, Andretta2017, Sanz-Forcada2008}, as can be seen
in the top panel of Fig.~\ref{temperature}. To quantify the relation between effective 
temperature and pEW(He IR), 
we computed Pearson's correlation coefficient using all valid measurements of absorption lines, which
yields a correlation coefficient of 0.75 with a p-value of $2.7\times 10^{-32}$.
Thus, we conclude that pEW(He IR) does depend on effective temperature.

Not surprisingly, a similar behaviour is identified when 
plotting the pEW(He IR) measurements 
as a function of spectral type instead of effective temperature, as can be
seen Fig.~\ref{temperature}, bottom panel. There we also provide the detection
fraction as a function of spectral type. No
valid detections were made for stars later than M5, whereas for the earlier M dwarfs, the valid detection
rates approach 80\%.   More specifically, 
among the very earliest M~dwarfs, the detection fraction of the \hei\,
line is close to $100$\,\%, and 
there is only one M0.0\,V and one M1.5\,V star (J04219+213/LP~415-17 and J15218+209/OT~Ser, respectively), where
we failed to detect the  He\,IR line.
Visual inspection of the spectrum of J04219+213/LP~415-17 suggests that the non-detection of the
\hei\ line is actually caused by noise artefacts, while
the spectrum of the young active star OT~Ser displays only a
shallow absorption line.
This star is the only one of its spectral sub-type exhibiting H$\alpha$ in emission and shows
at least two flares in the CARMENES time series. Not surprisingly, also
the individual spectra of the \hei\ lines show variability,
including absorption and emission in the \hei\, line.
We therefore argue that in the case of OT~Ser the \hei\, line is actually filled in in 
the averaged spectrum of this star, which hindered its formal detection.

Beyond the spectral sub-type M1.5\,V, the detection fraction of the \hei\ line decreases systematically. 
For sub-types M2.0\,V and M2.5\,V, six stars have non-detections of the \hei\, line, and two of them show 
artefacts of telluric airglow lines in the averaged spectrum that misled the fit.  Another two stars seem 
to show very shallow absorption lines, which are considered invalid by our automatic selection procedure.
The two remaining stars have quite flat spectra in the region of
the \hei\, line. The latter four stars are among the least active 
in the groups pertaining to their spectral sub-type as measured by the pEW(\ion{Ca}{ii} IRT) and
they show also no sign of activity in their H$\alpha$ lines.
Therefore, we argue that the \hei\, line is not filled in,
but is most likely to be intrinsically weaker than in the other stars of this spectral 
type.

At spectral type M4.0--M4.5\,V, the detection fraction has significantly declined.
Out of the 75 sample stars in this spectral type regime, 
we end with only 18 valid pEW(He IR) measurements. The 57 stars without valid pEW(He IR) measurements
span the whole range of activity levels seen in stars with this spectral sub-types.
Scrutinising the spectra of these stars with high activity levels as indicated by low
pEW(Ca IRT) values, we find that they are affected by variability, which effectively acts as a fill-in in the
average spectra. These active stars may have a detectable \hei\ line during pure quiescent states (which
we do not see because of averaging in the flaring phases as identified by looking at the variability in the
H$\alpha$ emission).
In contrast, the spectra of the inactive stars
really have shallow or flat \hei\ lines. 

Turning now to even later sub-types, among the 23 M5.0\,V stars in our sample, a valid measurement of 
the pEW(He IR) could only be obtained for the star J01019+541/G~218-020. This star
is a fast rotator with $\varv\sin(i) = 30$~km\,s$^{-1}$ measured by \citet{Reiners2017} and could be
associated with the Ursa Majoris moving group \citep{carmencita}.
The remaining 22 M5.0\,V sample stars with \hei\ non-detections have
a wide range of rotation periods and H$\alpha$ activity levels with the majority showing
H$\alpha$ in emission.
A contribution of the \hei\, line is revealed by its variability for many of these stars
when looking at the individual spectra. Since we have only snapshots rather than 
continuous observations it is hard to identify the reasons for temporal activity variations in these stars.
The H$\alpha$ line can reveal obvious flaring, but may remain inconclusive when observed
in the late decay phases of flares, when it may have returned to quiescent levels while other
chromospheric lines are still in emission.
Although periodic modulation by rotation or even cycles may contribute to the variability in chromospheric lines,
we consider flaring to be the most probable explanation for the mid M dwarfs, because it is a phenomenon
frequently observed for such objects.

Nonetheless, it appears that the \hei\ line is weak or absent
during the most inactive phases of these late M stars. These findings are consistent with a weakening of the line
towards later spectral types and a resulting decline in the detection fraction.

The vast majority of the M-type sample stars later than M4.0\,V shows H$\alpha$ in emission, demonstrating the
existence of chromospheres (and presumably coronae) in these objects.   This might suggest a
general filling in of the line as an explanation for the observed weakening of the line towards 
later spectral types. However, we argue against this hypothesis as follows:
we do not observe the \hei\,IR line going into emission during the
quiescent state, which is what we would expect to happen had there been any activity-driven fill-in for the line. 
We consider it unlikely that the \hei\, line is
fully filled in, but it is neither visible in absorption nor in emission during the quiescent state. 
However, it appears that the line can easily be observed in emission during flares, when
EUV fluxes are enhanced. Moreover, the densities are likely to be much higher and, therefore, collisions may also
contribute significantly to the line emission. Another explanation for the observation of an emission line 
is that the emitting material may be
located off-limb, dominating the line flux.

Therefore, it appears likely that for stars later than about M4, the conditions for the formation of the
\hei\, line are different in comparison to those for stars of an earlier type. This refers to the
upper chromosphere and lower transition region, where the \hei\, line is expected to originate
(in the case of the photo-ionisation and recombination process, the line is formed in the upper chromosphere, while collisional processes are expected to take place in the lower transition region for temperatures exceeding 20\,000 K
\citep{Andretta1997}).
Possible changes inhibiting
the formation of the \hei\, line encompass: ($i$) a lower density in the chromosphere so that
fewer atoms can interact with the photospheric light (but, in contrast, the H$\alpha$ emission in those stars
demonstrate relative high densities), ($ii$) a lower EUV radiation level, which leads to less \hei\,
ionisation and therefore fewer recombinations to the meta-stable lower level
of the \hei\,IR line (although high levels of X-ray radiation persist for later spectral type
stars in our sample, see Sect. \ref{X-ray}), ($iii$) a reduced geometrical thickness of the layer where the \hei\ line
is formed.  The latter possibility was demonstrated in simulations based on solar chromospheric models
by \citet{Avrett} which resulted in a reduced \hei\ line absorption for the same coronal illumination and chromospheric
densities.

Finally, the steep decrease in the detection fraction of the \hei\,IR line
takes place at those spectral types where M~dwarfs ought to become fully convective,
which is believed to happen around M3.5.  However, this concurrence might be coincidental and, thus,
proof of a physical link between these two phenomena remains elusive at this moment.

\subsection{Comparison to previous studies of the \hei\,IR line}\label{previous}

\citet{Patrick} have also recently used the CARMENES sample of spectra to study chromospheric
activity in various spectral lines.  \citet{Patrick}
use the spectral subtraction technique, which
relies on subtracting the spectrum of
an inactive reference star before any pEW measurement. This technique is especially
efficient for chromospheric lines that have a photospheric contribution to be
subtracted out, which is not the case for the \hei\,IR line.
For each spectral sub-type, \citet{Patrick} use the spectrum of the star with the longest
(known) rotation period as a reference. Therefore, these results are effectively 
`differential pEW' measurements that are defined with respect to the chosen reference star. 

\citet{Patrick} find that the differential pEW is about just as often
negative as positive for early M0--1\,V spectral types, with excess absorption occurring more often
in M1.5--4.0\,V type stars with respect to the template. 
Therefore \citet{Patrick} also measure some variation of their differential pEW(He IR) with spectral type.
Their
results are, nevertheless, not easily compared to ours
because the  dependence on spectral type is eliminated by the application of the spectral subtraction
of the reference star.

The \hei\ triplet has also been studied by \citet{Sanz-Forcada2008}, \citet{Andretta2017}, and
\citet{Dupree2018}.
In the upper panel of Fig.~\ref{temperature}, we additionally plot  
their measurements of the pEW(He IR) for twelve stars later than K0 (magenta diamonds).
Since these papers state spectral type or $B-V$ colour but no effective temperature, we made our conversions into temperature
based on  the relations provided by \citet{Mamajek}.
There are seven stars with effective temperatures below $4500$~K and, as shown in Fig.~\ref{temperature},
despite their low number, these complementary measurements show the same decrease in pEW(He IR)
towards lower effective temperature.

Moreover, these additional data suggest the existence of a `saturation regime' for \hei\ line 
absorption at a pEW of about $300$~m\AA\,, which appears to have been reached for the earliest M~dwarfs,
and extends towards higher effective temperatures.
Our largest pEW measurements are just below 0.3 \AA,  which is
well below the limit of 0.4 \AA\, for single active dwarfs, which \citet{Andretta1995} derived
theoretically based on model calculations for solar-type stars.

\subsection{Dependence of H$\alpha$ and \ion{Ca}{ii} IRT on effective temperature}\label{halpha_temperature_dependence}

\begin{figure*}
\begin{center}
\includegraphics[width=0.5\textwidth, clip]{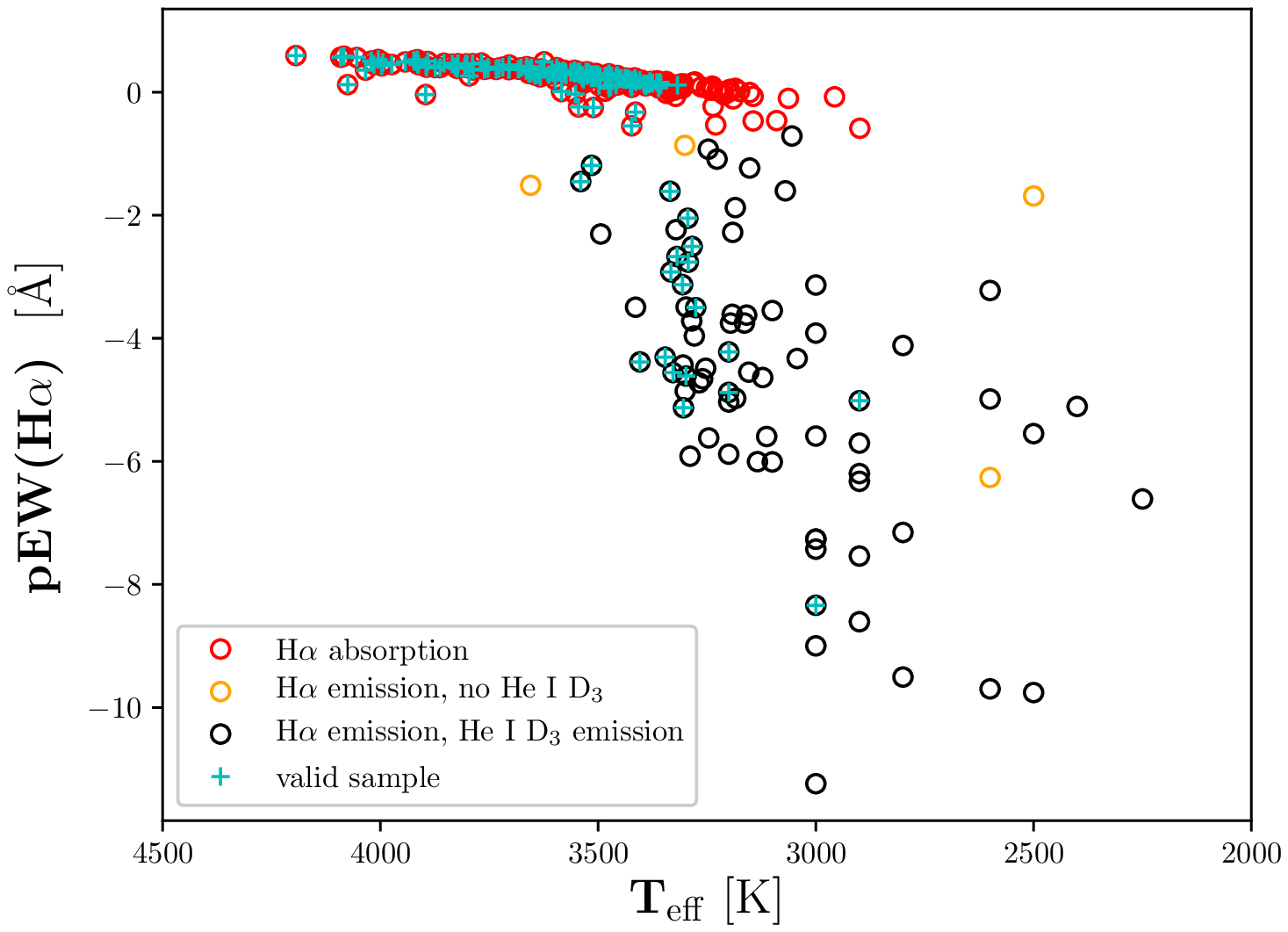}
\includegraphics[width=0.5\textwidth, clip]{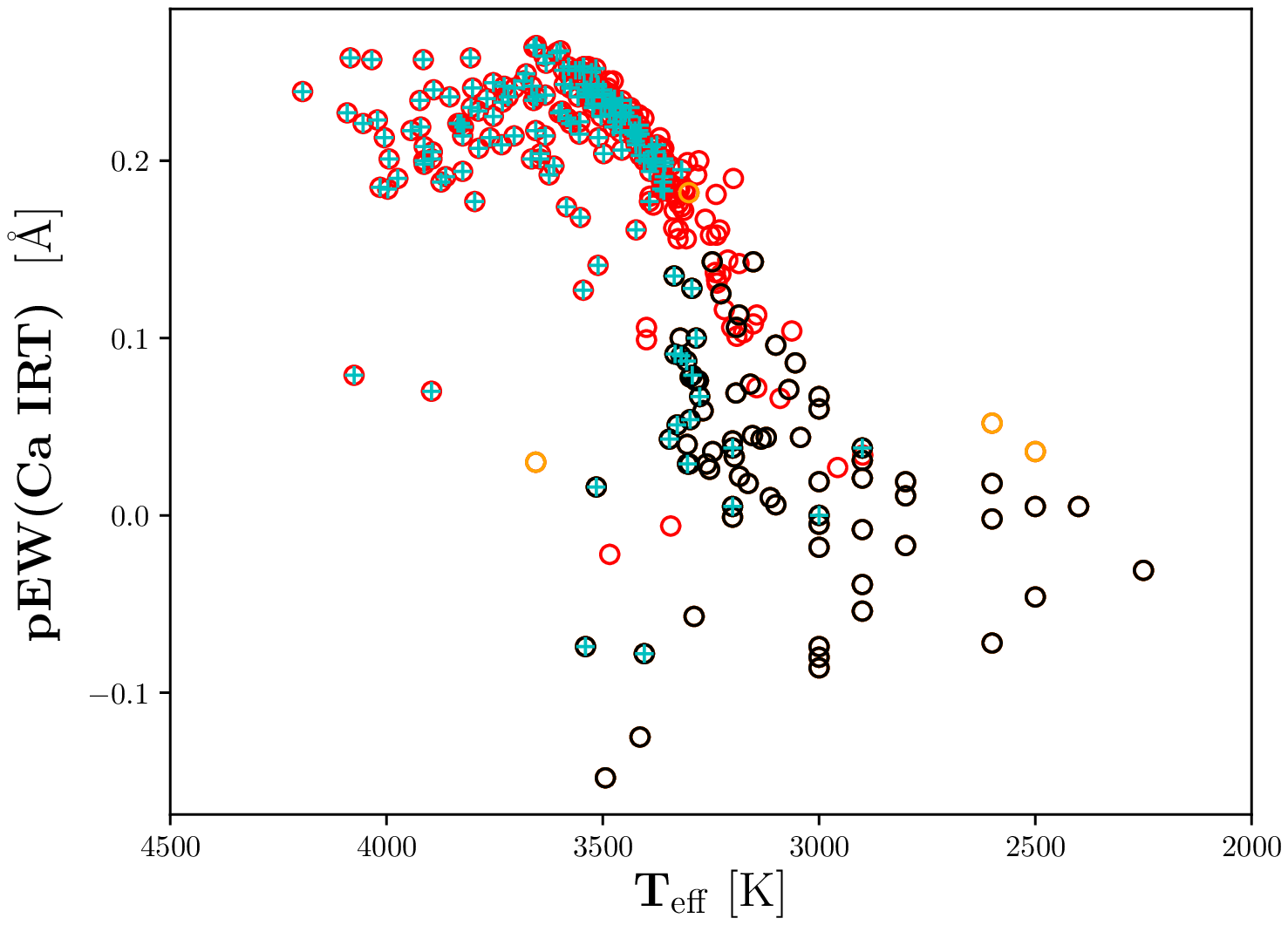}\\
\caption{\label{temperature_halpha} Relation between stellar effective temperature and pEW for the
  chromospheric H$\alpha$ (\emph{left}) and \ion{Ca}{ii} IRT lines (\emph{right}).
  Cyan crosses represent the valid sample as defined for the \hei\,IR line measurements.
  Red circles represent stars with H$\alpha$ in absorption. Black circles
  represent stars with both the H$\alpha$ and the \ion{He}{i}~D$_{3}$
  line in emission, and orange circles symbolise stars with spectra showing only H$\alpha$
  in emission.  }
\end{center}
\end{figure*}

In Fig.~\ref{temperature_halpha} we show 
the dependence of the pEW measured for the H$\alpha$ and \ion{Ca}{ii} IRT lines on photospheric
temperature. Since measuring pEW(H$\alpha$) and pEW(Ca IRT) in all sample stars is
straightforward, we show our measurements for the whole stellar sample. The valid
sample in terms of meaningful  \hei\,IR line measurements is indicated by cyan crosses. 
We further indicate whether H$\alpha$ is seen in absorption (red circles) or emission
(black circles). For most stars, H$\alpha$ seen in emission goes along with
\ion{He}{i}~D$_{3}$ seen in emission, and those stars where H$\alpha$ is seen in emission
without \ion{He}{i}~D$_{3}$ are indicated by the 
orange circles.   

The measured value of pEW(H$\alpha$) is frequently used to formally differentiate between active
stars with H$\alpha$ in emission and inactive stars. However,
the actual pEW(H$\alpha$) threshold
below which an H$\alpha$ line is considered to be a true emission line has varied between  
different authors:
\citet{Jeffers2018} adopt $-0.5$~\AA; \citet{West2011}  $-0.75$~\AA; and
\citet{Newton2017}  $-1.0$~\AA.
Here have applied a threshold of $-0.6$~\AA\ for the pEW(H$\alpha$) value to mark the
active stars in Fig.~\ref{temperature_halpha}.

Based on this criterion, all active stars in our sample
have effective temperatures cooler than about 3500~K, and with the exception of
two stars, all sample stars below 3000~K show H$\alpha$ in emission. 
The pEW(H$\alpha$) values of these active H$\alpha$ emitters form a widely spread-out cloud.
Nevertheless, there is a loose correlation between their pEW(H$\alpha$) and effective temperature with a
Pearson r-value of 0.5 and a p-value of $1.8\times10^{-6}$.
For the inactive dwarfs with H$\alpha$ seen in absorption there is a rather tight relation,
such that the pEW(H$\alpha$) values decrease
towards cooler stars, a result also found by \citet{Stauffer}, \citet{Newton2017}, and \citet{Jeffers2018}.
 
A similar behaviour is seen for the \ion{Ca}{ii} IRT line. For stars with H$\alpha$ in emission,
 the pEW(Ca IRT) values also show large scatter. For the more inactive stars, the pEW(Ca IRT)
declines with cooler effective temperature, which is most clearly
seen for stars with effective temperatures between about 3800~K and 2800~K. 
The \ion{Ca}{ii} IRT line goes into emission only for the most active stars.

\subsection{Relation between the \hei\,IR line and the H$\alpha$ and \ion{Ca}{ii}~IRT lines}

\begin{figure*}
\begin{center}
\includegraphics[width=0.5\textwidth, clip]{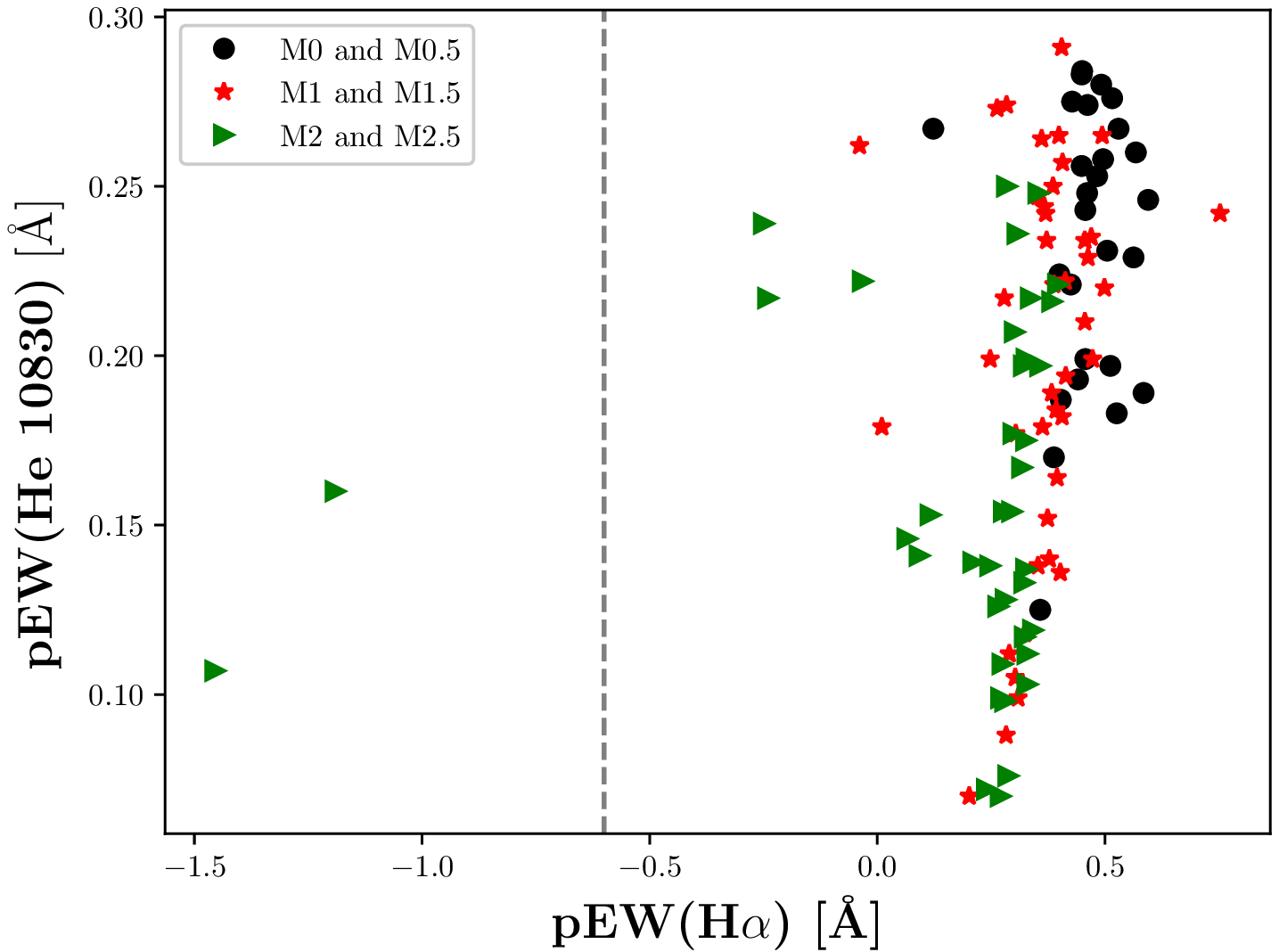}
\includegraphics[width=0.5\textwidth, clip]{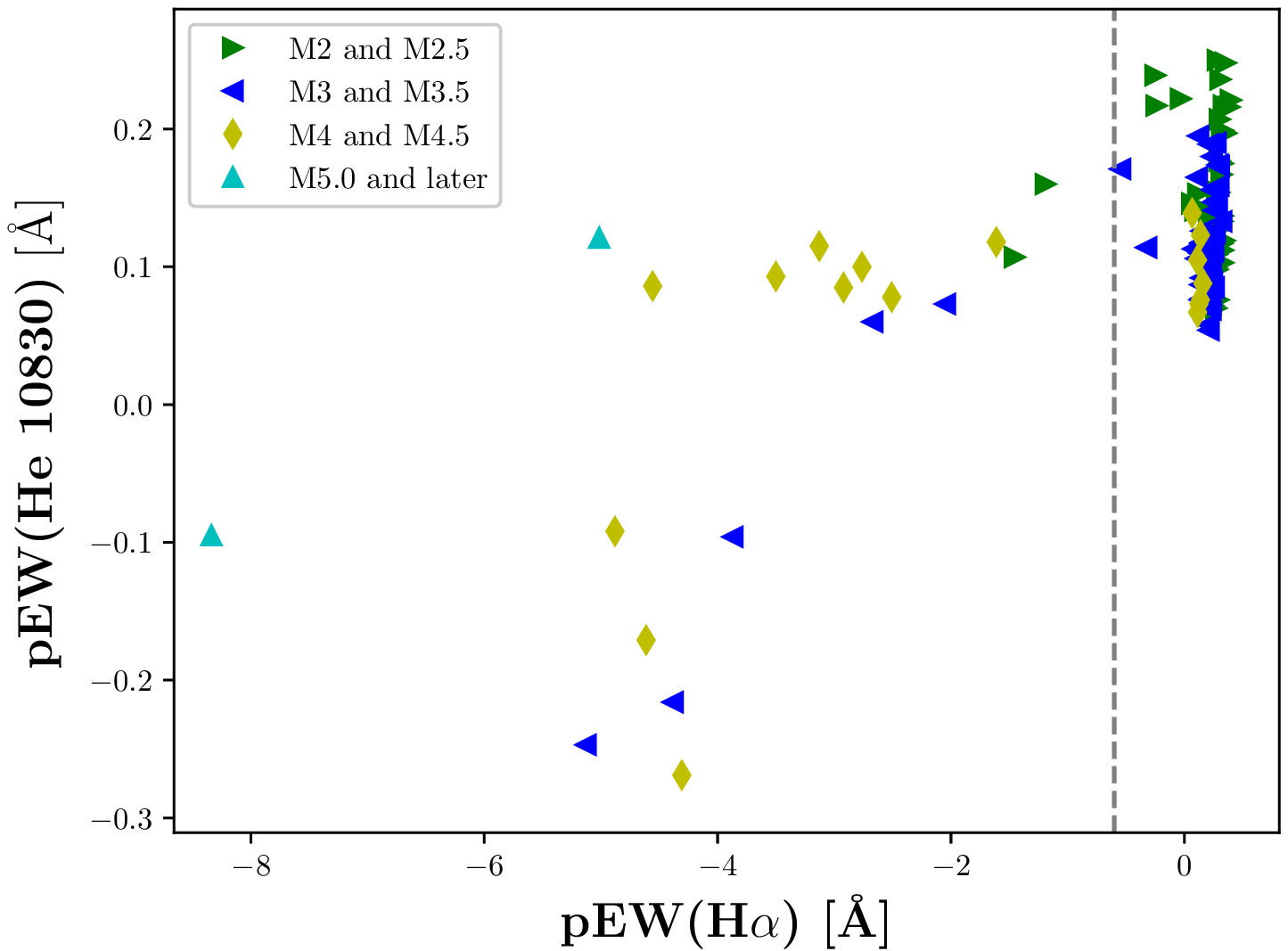}\\
\caption{\label{halphacorr} Valid measurements of pEW(He IR) as function of pEW(H$\alpha$)
  for different spectral types as given in legend. The dashed vertical line
  marks the dividing line between active (H$\alpha$ in emission) and inactive stars
  (H$\alpha$ in absorption) for both panels. The cloud of inactive late-type M dwarfs (\emph{right panel})
  connects smoothly to the cloud of the inactive early-type M dwarfs (\emph{left})
  shifting to lower maximum values
  for later spectral types. For better comparison, we show spectral type M2/M2.5 in both panels.}
\end{center}
\end{figure*}

We next turn to the relation between pEW(H$\alpha$) and pEW(He~IR) for individual spectral sub-types
shown in Fig.~\ref{halphacorr}. We note that  only stars with valid measurements of pEW(He IR) are 
considered here.   The left panel shows the early spectral sub-types, while the right panel shows the
  late sub-types. The Figure indicates that the H$\alpha$
  active stars demonstrate a  correlation of pEW(He~IR) to pEW(H$\alpha$), while the inactive
stars are grouped in a tight band. 
Looking at the left panel, for sub-types earlier than about M3 for the inactive stars,
  a nearly vertical band can be noted for each sub-type. When looking specifically at
  M1.0\, V stars, higher
values of pEW(He IR) are shown to correspond to higher values of pEW(H$\alpha$). This implies that both
lines become deeper absorption lines until a maximum value is reached. 
Then the trend reverses and both lines start to fill
in and finally go into emission (only for sub-types later than or equal to M2 is H$\alpha$ emission observed).
In summary, for increasing levels of activity
for inactive stars both the H$\alpha$ and the \hei\, IR line become deeper absorption lines
until a certain (sub-type dependent) value is reached, when both lines start to become more shallow again, that is, they fill in.

For spectral sub-type M4, it appears that pEW(H$\alpha$) decreases for increasing pEW(He~IR), indicating that the H$\alpha$ line has already started to fill in for the inactive stars. For  spectral
sub-types M5\,V and later, there are no inactive stars in the sample.
Generally, for the active stars, the \hei\,IR line appears to show a decrease in pEW for
decreasing pEW(H$\alpha$), indicating an increasing activity level. However,
filling in does not appear to occur in quiescence, but it is, rather, associated with
flares or, at least, episodes of enhanced activity.
Particularly when it comes to the spectral sub-types M4 and later,
the picture in Fig.~\ref{halphacorr} is likely dominated by flares for the lowest values
of both pEW(H$\alpha$) and pEW(He IR). This is in line with increasing flare duty cycles
from 0.02\% for early M dwarfs to 3\% for late M dwarfs found by \citet{Hilton2010}.
In fact, visual inspection suggests that all sample stars with H$\alpha$ in emission
(pEW(H$\alpha$) < -0.6 \AA)
show changing levels of activity during observation, which
do not average out and, therefore, `contaminate' the pEWs to some extent.
However, in contrast to the \hei\,IR line, the H$\alpha$ and \ion{Ca}{ii} IRT
(and also the \hei\, D$_{3}$ line)
can turn into emission lines not only for flares, but also during the quiescent state.

\begin{figure}
\begin{center}
\includegraphics[width=0.5\textwidth, clip]{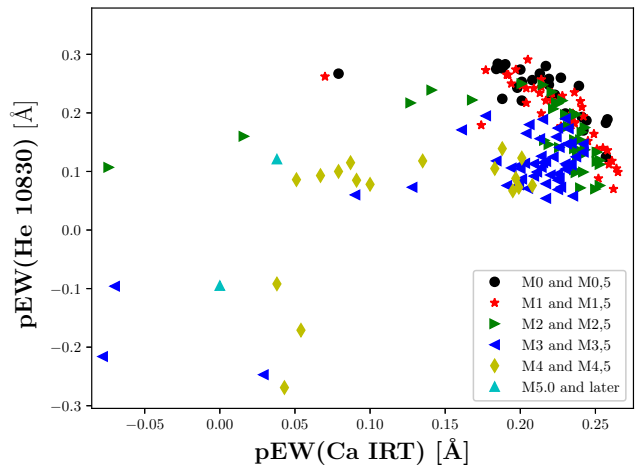}\\
\caption{\label{cairt_ew} Valid pEWs of \hei\,IR line
  as function of pEW(Ca IRT)
  for different spectral types as given in legend.}
\end{center}
\end{figure}

The \ion{Ca}{ii} IRT lines are also frequently used as activity indicators \citep{Mittag2017, Johannes}.
In Fig.~\ref{cairt_ew} we show the analogous relation between pEW(Ca IRT) and pEW(He IR).
Since the \ion{Ca}{ii}~IRT line only fills in for higher levels of
activity, the relation between the two quantities in the low-activity regime
is more pronounced when compared to the case of pEW(H$\alpha$).

\subsection{Relation between the \hei\, D$_{3}$ and IR lines}

\begin{figure}
\begin{center}
\includegraphics[width=0.5\textwidth, clip]{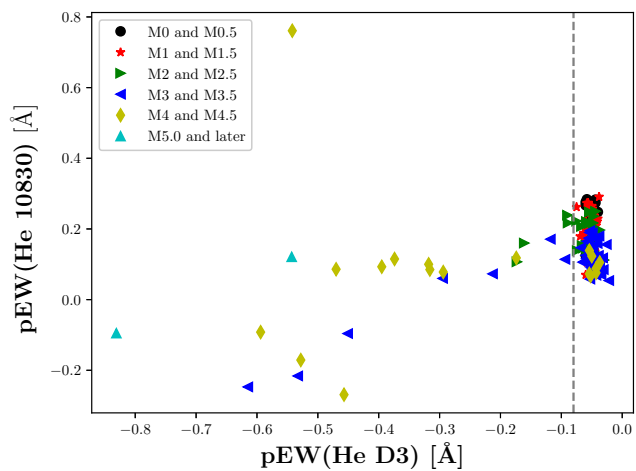}
\caption{\label{he5877corr} Valid (according to our selection criterion) pEW(He IR) as function of pEW(\hei\, D$_{3}$)
  for  different spectral types given in legend. The grey dashed vertical
  line indicates
  the threshold for the line in emission for lower values or non-existent
  for larger values.}
\end{center}
\end{figure}

Since the lower level of the \hei\,D$_{3}$ line is the upper level of the \hei\,IR line,
simultaneous measurements of both lines can provide additional diagnostics. This is the method used, for example, by \citet{Andretta2017} to infer filling factors for active regions of solar-like stars by
comparison to solar chromospheric models.
In Fig.~\ref{he5877corr}, we show the relation between the measured pEWs for the
\hei\, D$_{3}$ line at $5877$~\AA\ and the \hei\,IR line.
Among the inactive stars in our sample, we neither detect the \hei\, D$_{3}$ line in
absorption nor emission. The measurements by \citet{Andretta2017} suggest that the
\hei\, D$_{3}$ line is very weak and, therefore, difficult to detect on top of the molecular pseudo-continuum
of our stars (see Fig. \ref{hed3specobs}).
Moreover, due to  normalisation effects, we measured a slightly negative pEW(He D$_{3}$) for
inactive stars (see Fig.~\ref{he5877corr}). Therefore, we opted for a threshold
of pEW(He D$_{3}$) < $-0.08$~\AA\ for the acceptance of a measurement as an emission line.
For the 24 stars for which a non-zero value for pEW(He D$_{3}$) could be
measured, it is correlated to pEW(He IR). Moreover, pEW(He D$_{3}$) is tightly correlated with
pEW(H$\alpha$). Consequently,
these active stars show a spectral type-dependent correlation, which is almost identical
to that seen for the H$\alpha$ line.

In stars where  the \hei\, D$_{3}$ line is seen in emission, the \hei\,IR line
can be in absorption (13 stars) or in emission (7 stars).
For the latter seven stars, all show $L_{\rm X}/L_{\rm bol} > -3.5$ and effective temperatures below 3550~K.
Furthermore, for all sample stars with the  \hei\, D$_{3}$ line in emission and the \hei\,IR line
detected, H$\alpha$ is also in emission. 
Visual inspection reveals that all of these 
stars undergo temporary enhancements in their H$\alpha$ emission, which we attribute to flaring activity.
The \hei\,IR line is, therefore, also affected by flaring, and more severely or more often for the stars where
it is seen in emission. The \hei\,IR line is observed during the quiescent state
as an absorption line or not observed at all
in these stars. By contrast, the \hei\, D$_{3}$ line can also be observed in emission  during the quiescent state. 

Unlike the \hei\ IR line,
the \hei\, D$_{3}$ line
tends to go
into emission for late spectral types, which we interpret as a (quasi-)~continuum effect.
PHOENIX spectra demonstrate that for mid M dwarfs the continuum around the
\hei\, D$_{3}$ is a factor of 100 
lower than at the wavelength of the \hei\, IR line, while it is
comparable for M0 stars. A detection of the \hei\, D$_{3}$ line against the weak background in
late-type M~dwarfs is, therefore, less challenging than in early-type stars.
If the line is truly absent in the inactive dwarfs or is merely undetectable, remains unknown.
Yet it seems  that a minimum level of activity is needed to drive the line
into notable emission, and this level coincides with the parameter beyond which H$\alpha$ goes into 
emission as well.

\citet{Andretta2017} also found a correlation between the \hei\, D$_{3}$ and the \hei\,IR lines
for earlier-type stars. However, the authors find \hei\, D$_{3}$ in absorption,
which we never observe for the CARMENES M dwarfs, and they derive a positive correlation 
between pEW(He IR) and pEW(He D$_{3}$),
measuring values from 0.003 to 0.07~\AA\ for the latter. Our study is
not sensitive enough to measure such small pEWs in the presence of the surrounding photospheric
molecular lines.
The sample of \citet{Andretta2017} reaches down to a spectral type of about K3, so that there is
a gap between the coolest stars in their sample and the hottest ones in our
sample. Their latest-type stars show the lowest pEW(He D$_{3}$) values, which is consistent with
our results.
\citet{Houdebine} use a spectral subtraction
technique to study the \hei\, D$_{3}$ line in 37 M1 stars and find measurable absorption for one
M1 dwarf (H$\alpha$ in absorption) and eight M1e dwarfs (H$\alpha$ in emission),
but they do not find emission (just as we also found no emission) in these early M dwarfs. The stars 
with the strongest 
absorption in their sample are, unfortunately, not included in our sample.

\subsection{Relation between  chromospheric helium lines and  X-ray emission}\label{X-ray}

\begin{figure*}
\begin{center}
  \includegraphics[width=0.5\textwidth, clip]{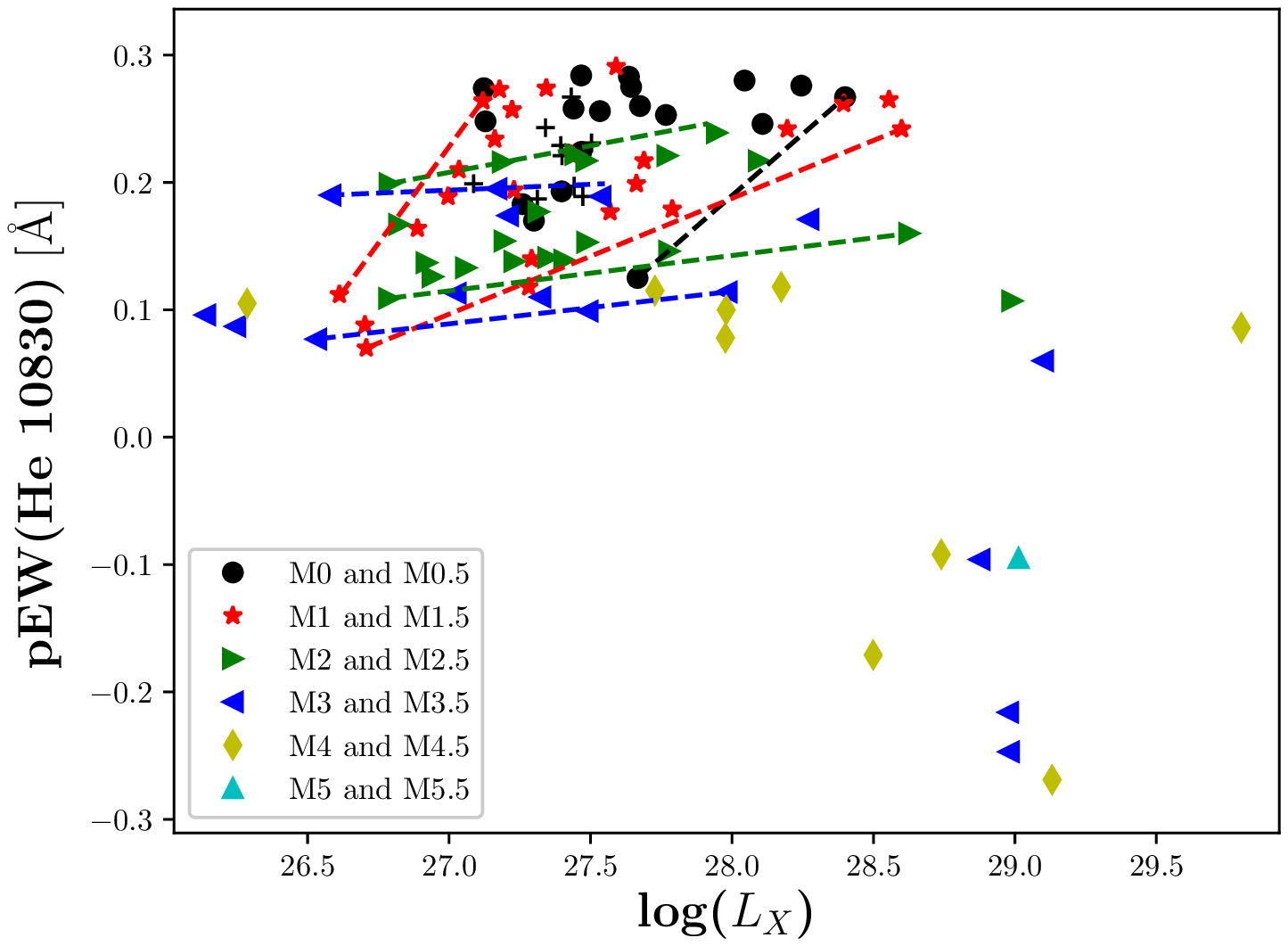}
  \includegraphics[width=0.5\textwidth, clip]{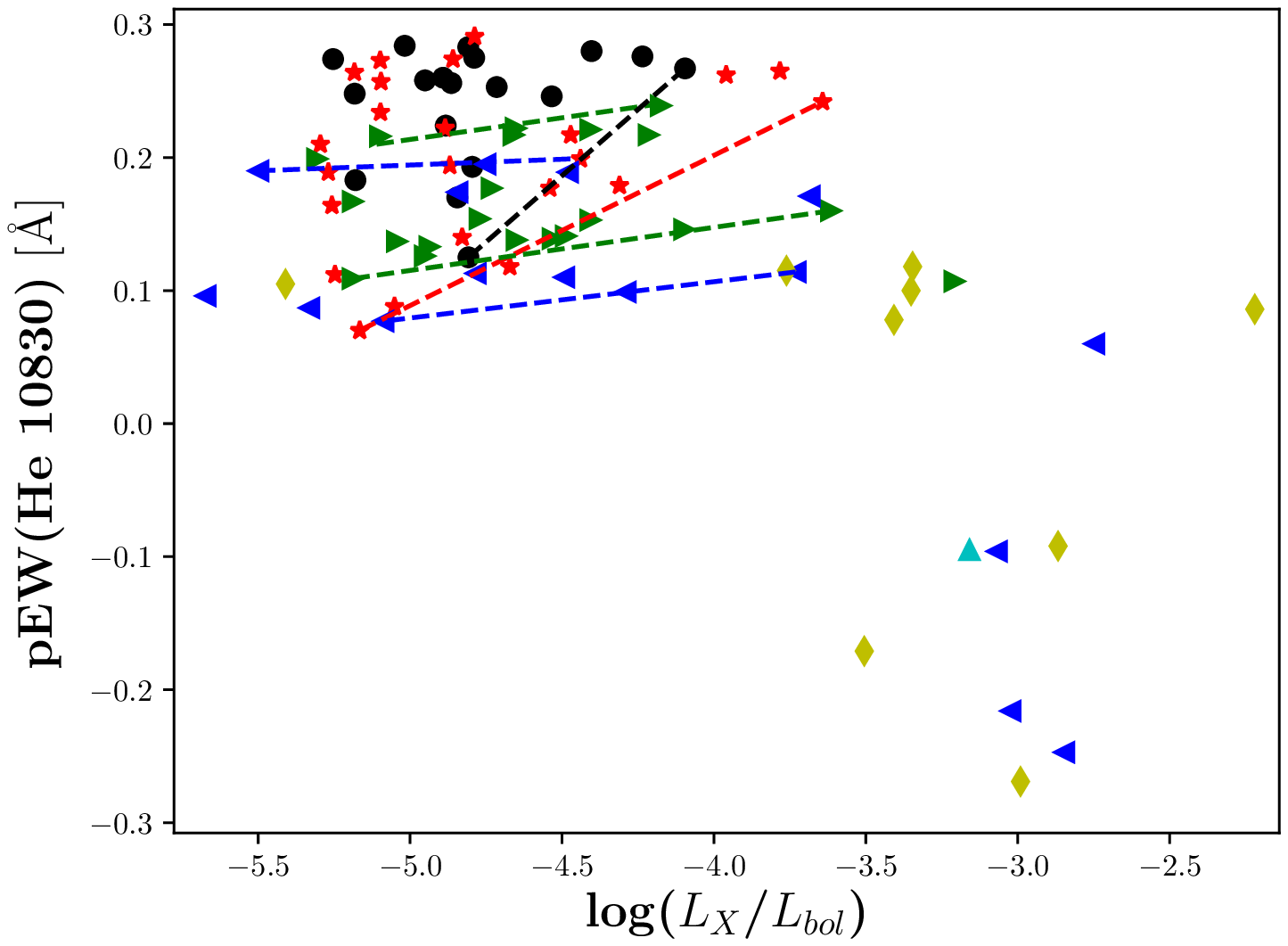}\\
  \caption{\label{lxlbol} Measurement of pEW of \hei\,IR 
     as function of $L_{X}$ (\emph{left panel}) and of  $L_{X}/L_{bol}$ (\emph{right panel}). 
Symbols
represent the pEW(He IR) of the valid sample with known X-ray fluxes, and colours decode spectral
sub-type as in the legend in the left panel. In both panels, dashed lines with colours according to sub-type
indicate hypothesise lower and upper envelopes where recognisable, while for the other spectral sub-types only
lower envelopes are indicated.
}
\end{center}
\end{figure*}

Several authors, such as \citet{Zarro1986}, \citet{Takeda2011}, and \citet{Smith2016}, 
have reported a correlation between the EW of the \hei\,IR line
and the X-ray luminosity ($L_{\rm X}$) scaled by the bolometric luminosity ($L_{\rm bol}$) for stars of
spectral type K and earlier, and argue that this finding is in line with the photo-ionisation and recombination mechanism: 
if more soft X-ray and EUV photons are present to ionise helium,  more electrons eventually
end up in the meta-stable $^3S_1$ ground state of the \hei\,IR triplet lines,
which then carry out more absorptions that cause a deeper absorption line.
While \citet{Sanz-Forcada2008}
find no correlation between X-ray flux and pEW(He IR) for highly active earlier-type
dwarfs, they do find it for giants and low-activity dwarfs. They attribute
this discrepancy to different levels of population mechanisms. In the thinner atmospheres of giants, the photo-ionisation and recombination mechanism 
dominates, while collisions may prove more significant
in the far denser dwarf atmospheres.

In Fig.~\ref{lxlbol} we show the distributions of pEW(He IR) as a function of $\log\left(L_{\rm X}\right)$
in the left panel and of $\log\left(L_{\rm X}/L_{\rm bol}\right)$ in the right panel
for our sample stars. 
The $L_{\rm X}$ values were derived from X-ray fluxes given in the Carmencita database
\citep{carmencita}, which were
measured mainly within the \textit{ROSAT} all-sky survey in the 0.1 -- 2.4 keV energy band \citep{Voges1999}.
For 154 stars, we actually obtain $L_{\rm X}$ detections
and upper limits on the X-ray flux for the remaining stars based on the \textit{ROSAT} survey, but
we find that including \textit{ROSAT} upper limits does not contribute any significant
information relevant to the analysis carried out here.
The resulting $\log\left(L_{\rm X}/L_{\rm bol}\right)$ values range from $-5.7$ to $-2.2$ for the
  stars in our sample, while $\log\left(L_{\rm X}\right)$ values range from 26.1 to nearly 30.0.
  A number of stars in our sample exhibit $\log\left(L_{\rm X}/L_{\rm bol}\right)$ values in excess of the
saturation limit of about $-3$ \citep[see, e.\,g.][]{Pizzolato}, suggesting that
X-ray flaring probably affected the measurement, especially in the most
extreme cases. 
For reference, we also show the $L_{\rm X}/L_{\rm bol}$ values
as a function of effective temperature in Fig.~\ref{tefflxlbol}.
Obviously, none of these X-ray measurements have been carried out simultaneously to our CARMENES data.  As a result,
this procedure may lead to comparisons between measurements at different activity levels, in
particular, flaring X-ray luminosity and quiescent pEW(He IR) values may be compared
or vice versa.

In the following, we focus on stars whose measurements are unaffected by
  flaring, namely, those for which we obtained $\log\left(L_{\rm X}/L_{\rm bol}\right) < -3$
  and absorption in the \hei~IR line, that is, a positive value for pEW(He~IR).
  Here we note that the population of the diagram at the low X-ray luminosity end is
  limited by the sensitivity of X-ray measurements.
  In particular, there is a lack of
  stars with measurements of $\log\left(L_{\rm X}/L_{\rm bol}\right) < -5.5$  which would support a relation according to the studies by
  \citet{Sanz-Forcada2008} or \citet{Zarro1986}.
  In Fig.~\ref{lxlbol}, we
  indicate estimates for the lower and upper envelopes of the distribution for each spectral sub-type.
  Neither distribution shown in Fig.~\ref{lxlbol} shows a clear correlation between pEW(He~IR)
  and X-ray properties. What the diagram does show is that for comparatively low activity levels of
  $\log\left(L_{\rm X}/L_{\rm bol}\right) = -5$, the whole range of pEW(He~IR) can be observed. This
  is particularly pronounced for the M1\,V spectral types.  
  For the early types M0\,V and M1\,V,
  the distributions further suggest a dividing line indicating that low values of pEW(He~IR)
  are not observed at high activity levels. This part of the diagram is more complete in the sense that
  the majority of active early type stars have X-ray detections. 
  For the spectral types later than M3.0\,V, the tendency cannot be
  discerned and \hei~IR and X-ray properties appear essentially uncorrelated.   

The \hei\,IR line is
only seen in emission
in stars which are also very active by X-ray standards ($\log\left(L_{\rm X}/L_{\rm bol}\right) \geq 3.5$).
There is a tendency for more X-ray luminous stars to also show stronger emission in
the helium line. The same is true for the \hei\, D$_{3}$ line. We find a correlation between pEW(He D$_{3}$) and $\log\left(L_{\rm X}/L_{\rm bol}\right)$
for these active stars with a Pearson correlation coefficient of $-0.44$ and a p-value
of $1.5\cdot 10^{-5}$. Similarly, the Pearson correlation coefficient between pEW(He IR) and $\log\left(L_{\rm X}/L_{\rm bol}\right)$
is $-0.24$ with a p-value of 0.02. Therefore, stronger X-ray emission tends to be associated with stronger emission in the helium lines. 
However, the measurements may be affected by X-ray flaring in this regime, especially in the most
extreme cases, and we believe that this is also the case for the measurement of the
helium lines. This is consistent with
these stars also acting as highly active H$\alpha$ emitters. Any correlation in this high-activity regime is, therefore, probably affected by flaring.

\section{Summary and conclusions}

We present a comprehensive study of the \hei\,IR line in the co-added CARMENES 
spectra of a sample of 319
M~dwarfs ranging from M0.0\,V to M9.0\,V with various levels of activity.
Since the surroundings
of the \hei\ line contain a number of unidentified spectral features, which are not
satisfactorily reproduced by synthetic spectra covering the range of M-type stars, 
we model the immediate spectral
environment of the \hei\,IR line using an empirical model consisting of four Voigt components, 
which provides adequate pseudo equivalent widths of the line.  Upon applying a conservative approach, we
obtained 181 valid measurements in our sample of 319 stars.   The detection rate of
the \hei\,IR line strongly depends on spectral sub-type.
Our results are consistent with the hypothesis that the \hei\ line is always present in the 
earliest-type M dwarfs, while the detection fraction continuously decreases with spectral sub-type and, 
for stars later than about M5, the \hei\ line is not detectable.

The observed strength of the \hei\,IR line depends on the effective temperature of the
star. In particular, it is strongest in early-type M~dwarfs where we find maximum pEW values of
around $300$~m\AA. The absorption strength of the line declines towards later spectral types. 
The line is usually seen in absorption with a few exceptions, which we attribute to flaring activity.
At least in these latter cases, the \hei\,IR line can exhibit prominent temporal variability, a detailed
discussion of which is beyond the scope of this paper.
In contrast, the \hei\, D$_{3}$ line is undetectable in the spectra of inactive stars,
while it is seen in emission in active stars regardless of spectral sub-type. We could not
establish the presence of any other \hei\, line in the CARMENES spectral range in our averaged
spectra.

We interpret the observed decline in pEW(He IR) with effective temperature in inactive stars 
without a detectable continuation into emission as evidence for a true, physical disappearance of the line as opposed to
an increasing level of fill-in caused, for example,  by collisional excitation.
If the line were collisionally controlled, rising activity levels would be expected 
to raise the population of the upper level, entailing first a fill-in in the line before driving it into emission. 
However, in our sample the weakest observed
\hei\,IR line absorption seems to correspond to the most inactive stars as measured by
\ion{Ca}{ii} IRT for the sub-types earlier than about M3.0\,V.


We studied the relation between pEW(He~IR) and the stellar X-ray properties. No
clear correlation between the \hei\,IR absorption and the X-ray emission is observed. Our data show that the
maximal level of \hei\,IR absorption can already be
reached at low activity levels of $\log\left(L_{\rm X}/L_{\rm bol}\right) = -5$. Moreover, we find no
early-type M~dwarfs in our sample that would show weak absorption in the \hei\,IR line at high activity levels.
Emission in the \hei\,IR line is only observed during flares in our sample.

Our results are consistent with the role of photo-ionisation and recombination mechanism
  as the main driver of \hei\,IR line formation.
  Although we do not detect the line in stars later than M5\,V, our study shows that the \hei\,IR
  triplet lines are a\ ubiquitous feature
in earlier-type M~dwarfs, which carries great diagnostic weight for future chromospheric structure and 
activity studies.

\begin{acknowledgements}
  B.~F. acknowledges funding by the DFG under Schm \mbox{1032/69-1}.
  CARMENES is an instrument for the Centro Astron\'omico Hispano-Alem\'an de
  Calar Alto (CAHA, Almer\'{\i}a, Spain). 
  CARMENES is funded by the German Max-Planck-Gesellschaft (MPG), 
  the Spanish Consejo Superior de Investigaciones Cient\'{\i}ficas (CSIC),
  the European Union through FEDER/ERF FICTS-2011-02 funds, 
  and the members of the CARMENES Consortium 
  (Max-Planck-Institut f\"ur Astronomie,
  Instituto de Astrof\'{\i}sica de Andaluc\'{\i}a,
  Landessternwarte K\"onigstuhl,
  Institut de Ci\`encies de l'Espai,
  Institut f\"ur Astrophysik G\"ottingen,
  Universidad Complutense de Madrid,
  Th\"uringer Landessternwarte Tautenburg,
  Instituto de Astrof\'{\i}sica de Canarias,
  Hamburger Sternwarte,
  Centro de Astrobiolog\'{\i}a and
  Centro Astron\'omico Hispano-Alem\'an), 
  with additional contributions by the Spanish Ministry of Economy, 
  the German Science Foundation through the Major Research Instrumentation 
    Programme and DFG Research Unit FOR2544 ``Blue Planets around Red Stars'', 
  the Klaus Tschira Stiftung, 
  the states of Baden-W\"urttemberg and Niedersachsen, 
  and by the Junta de Andaluc\'{\i}a.
  Based on data from the CARMENES data archive at CAB (INTA-CSIC).
  We acknowledge financial support from the Agencia Estatal de Investigaci\'on
  of the Ministerio de Ciencia, Innovaci\'on y Universidades and the European
  FEDER/ERF funds through projects AYA2015-69350-C3-2-P, ESP2016-80435-C2-1-R,
  AYA2016-79425-C3-1/2/3-P, ESP2017-87676-C5-1-R, and the Centre of Excellence ''Severo
  Ochoa'' and ''Mar\'ia de Maeztu'' awards to the Instituto de Astrof\'isica de
  Canarias (SEV-2015-0548), Instituto de Astrof\'isica de Andaluc\'ia
  (SEV-2017-0709), and Centro de Astrobiolog\'ia (MDM-2017-0737), and the
  Generalitat de Catalunya/CERCA programme''.
\end{acknowledgements}

\bibliographystyle{aa}
\bibliography{papers}

\appendix

\section{Examples of other chromospheric lines in the spectra}\label{otherhelium}
\begin{figure}[h]
\begin{center}
\includegraphics[width=0.5\textwidth, clip]{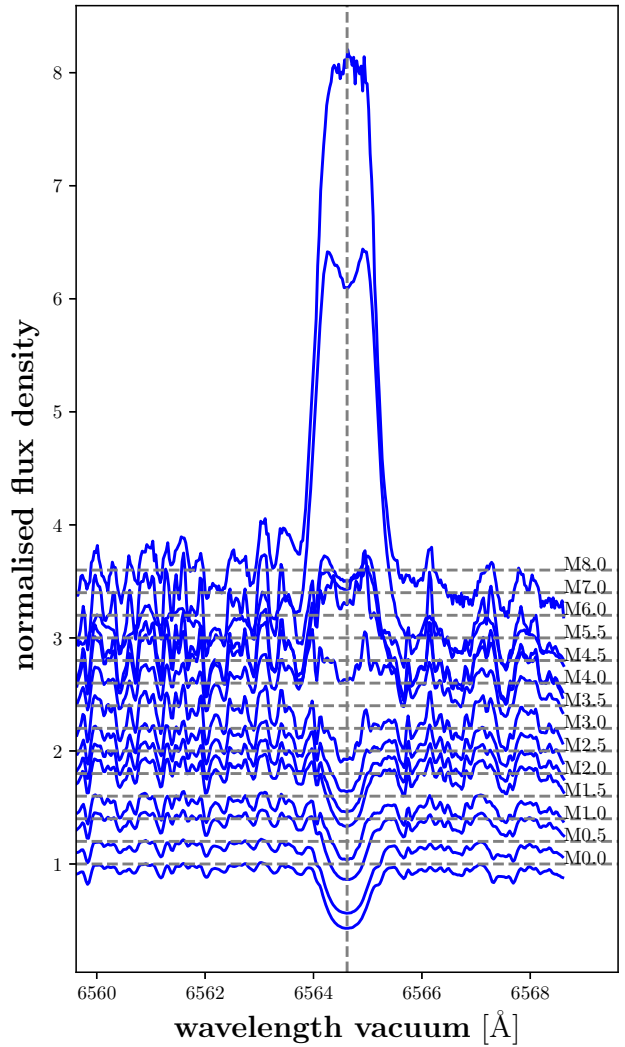}
\caption{\label{halphaspecobs} Spectral-type series for region around
  H$\alpha$.
  Each normalised spectrum (blue line) is offset by 0.2 in flux density for clarity purposes.
  The continuum for each spectrum is marked with a dashed grey line.
  The position of the line centre is marked with a grey dashed
  vertical line. While early M dwarfs are typically inactive (showing H$\alpha$
  in absorption), mid- to late-type M dwarfs are typically active, with very variable
  H$\alpha$ emission.
The shown stars are (same as in Fig. \ref{earlyandlateMdwarf}):
  M0.0\,V: J03463+262/HD~23453.
  M0.5\,V: J02222+478/BD+47~612.
  M1.0\,V: J00051+457/GJ~2.
  M1.5\,V: J02123+035/BD+02~348.
  M2.0\,V: J01013+613/GJ~47.
  M2.5\,V: J00389+306/Wolf~1056.
  M3.0\,V: J02015+637/G~244-047.
  M3.5\,V: J12479+097/Wolf~437.
  M4.0\,V: J01339-176/LP~768-113.
  M4.5\,V: J01125-169/YZ~Cet.
  M5.5\,V: J00067-075/GJ~1002.
  M6.0\,V: J14321+081/LP~560-035 
  M7.0\,V: J02530+168/Teegarden's star.
  M8.0\,V: J19169+051S/vB10.}
\end{center}
\end{figure}

\begin{figure}[h]
\begin{center}
\includegraphics[width=0.5\textwidth, clip]{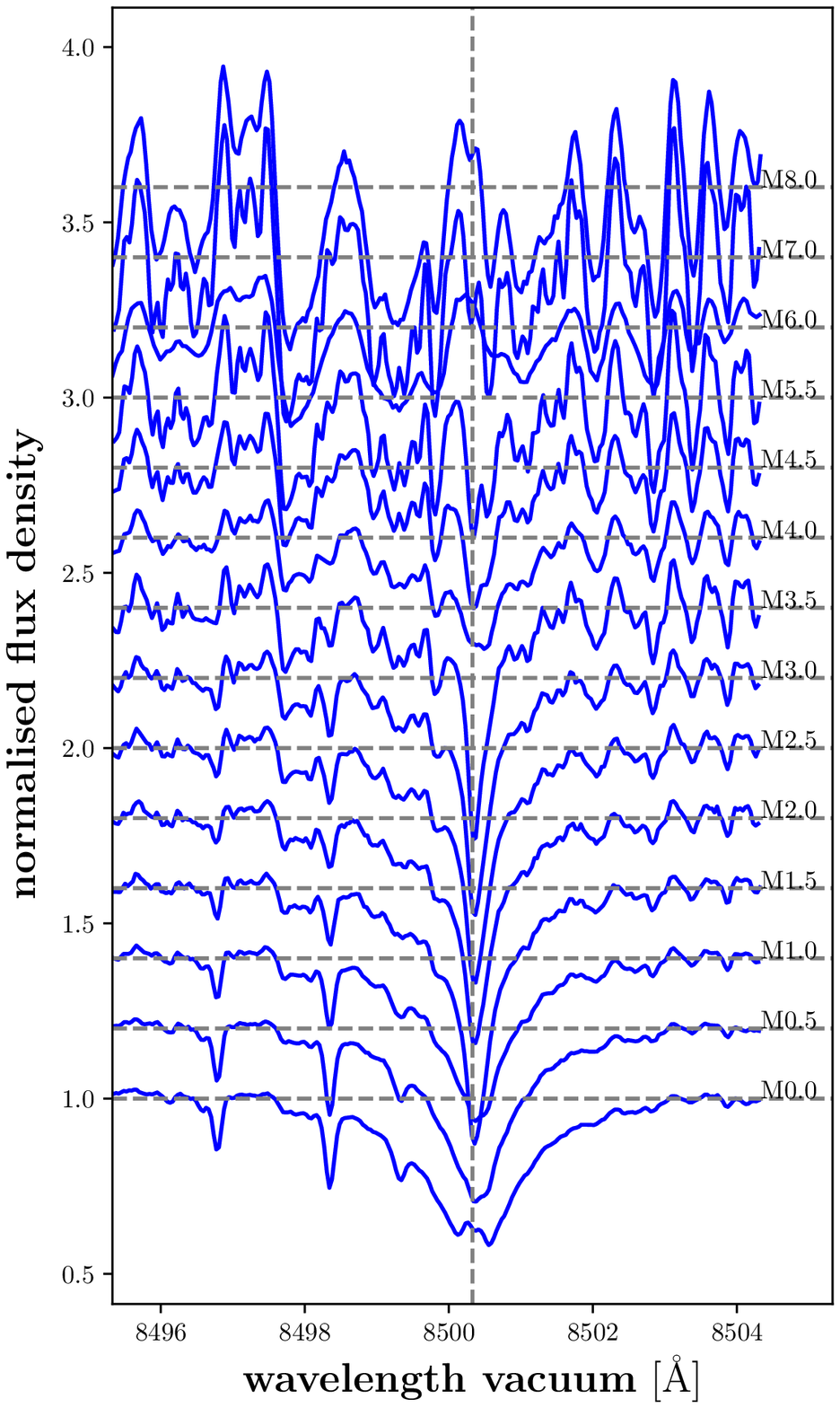}
\caption{\label{cairtspecobs} Spectral-type series for region around
  bluest line of \ion{Ca}{ii} IRT.
  Each normalised spectrum (blue lines) is offset by 0.2 in flux density for clarity purposes.
  The background for each spectrum is marked with a dashed grey line.
  The position of the line centre is marked with a grey dashed
  vertical line. 
The shown stars are the same as in Figs. \ref{earlyandlateMdwarf} and \ref{halphaspecobs}.
}
\end{center}
\end{figure}

\begin{figure}[h]
\begin{center}
\includegraphics[width=0.5\textwidth, clip]{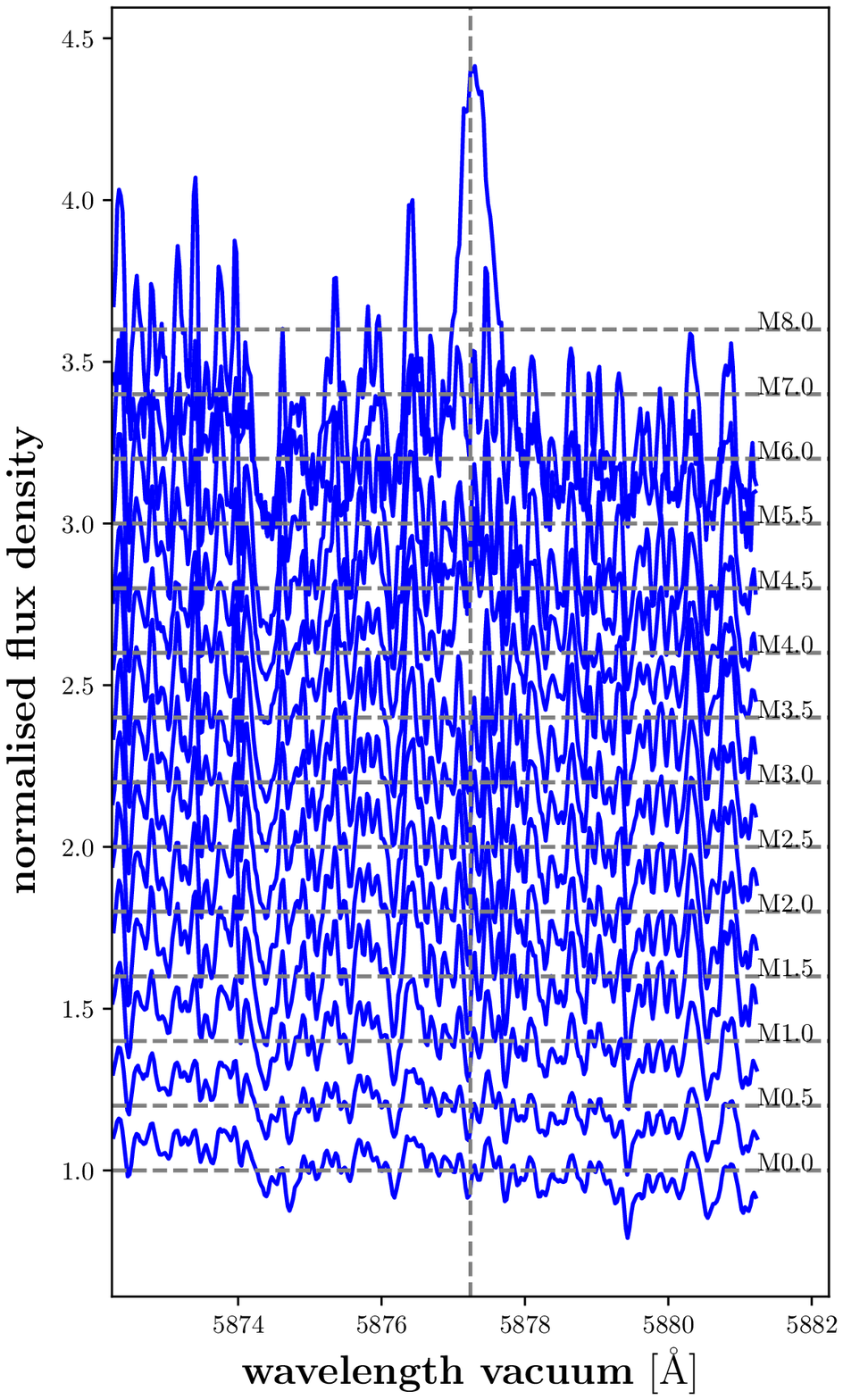}
\caption{\label{hed3specobs} Spectral-type series for  region around
   \hei\,D$_{3}$ line.
  Each normalised spectrum (blue line) is offset by 0.2 in flux density for clarity purposes.
  The background for each spectrum is marked with a dashed grey line.
  The position of the line centre is marked with a grey dashed
  vertical line. 
The stars shown are the same as in Figs.~\ref{earlyandlateMdwarf} and \ref{halphaspecobs}.
}
\end{center}
\end{figure}

Here we provide some representative examples of the other chromospheric lines used
as a comparison for the \hei\, IR line measurements. In Fig. \ref{halphaspecobs}
we show the same spectral-type series as in Fig. \ref{earlyandlateMdwarf}
but for the wavelength region around H$\alpha$. In Figs. \ref{cairtspecobs}
and \ref{hed3specobs} we show the same for the bluest \ion{Ca}{ii} IRT and the
\hei\,D$_{3}$ line. The slope in the continuum for the \hei\,D$_{3}$ line
is caused by the vicinity of the \ion{Na}{i} D lines, which have broad absorption
wings.

\begin{figure}[h!]
\begin{center}
\includegraphics[width=0.5\textwidth, clip]{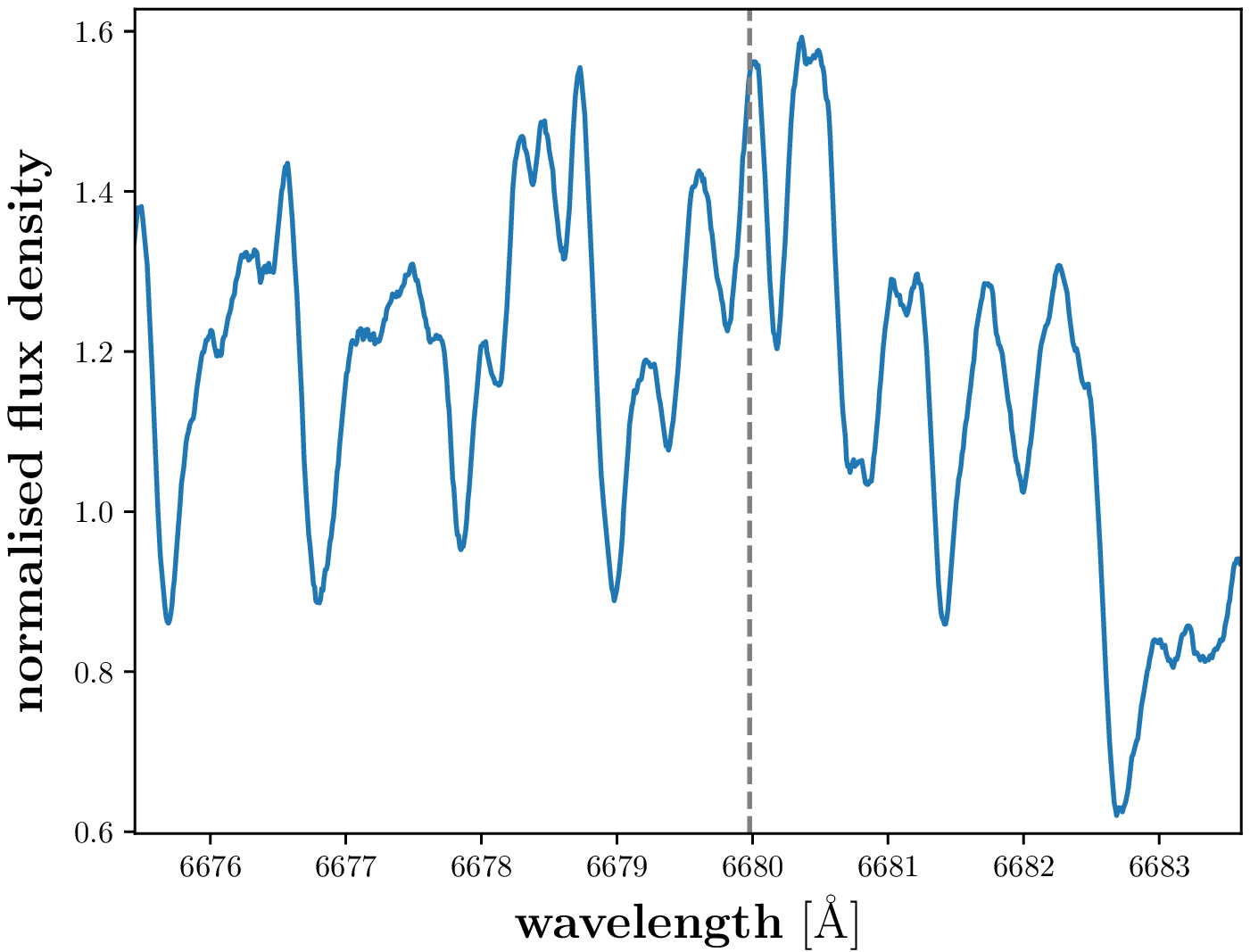}\\
\includegraphics[width=0.5\textwidth, clip]{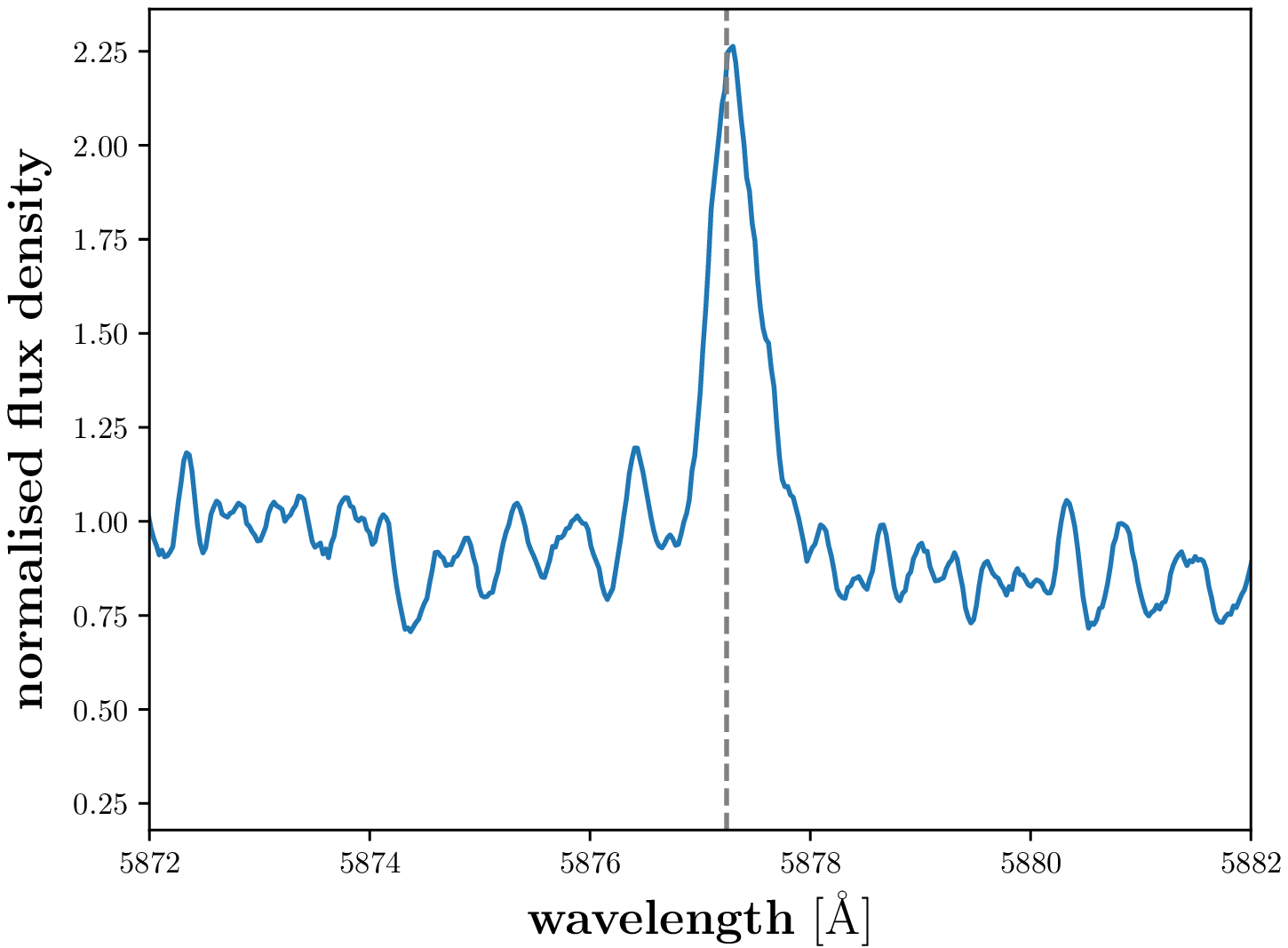}\\
\caption{\label{he6678} Examples of  spectral regions around  \hei\, line at 6679~\AA\,
  (\emph{top})
   and  \hei\, D$_{3}$ line at 5877 \AA\, (\emph{bottom}) for J07446+035/YZ~CMi.
  Positions of the two lines are marked with grey dashed
  vertical lines. }

\end{center}
\end{figure}

We want to illustrate our search for other \hei\, lines in the optical
by using as an example the very active M4.5\,V star J07446+035/YZ~CMi.
We show in Fig. \ref{he6678}, the \hei\, D$_{3}$ line which is in emission.
The slight asymmetry to the red
side is caused by another component of the line. On the other hand, the \hei\, line
at 6678 \AA\, is, at best, ambiguous, but most likely hidden in the molecular lines.

This is a typical example of a mid M dwarf, where H$\alpha$ and \hei\, D$_{3}$ are emission
lines, but other \hei\, lines in the optical are not seen during quiescent
state. For more early-type M dwarfs (which are usually inactive) the \hei\, D$_{3}$ is absent as well.

\section{X-ray luminosity compared to effective temperature}\label{appendixc}

\begin{figure}[h]
\begin{center}
\includegraphics[width=0.5\textwidth, clip]{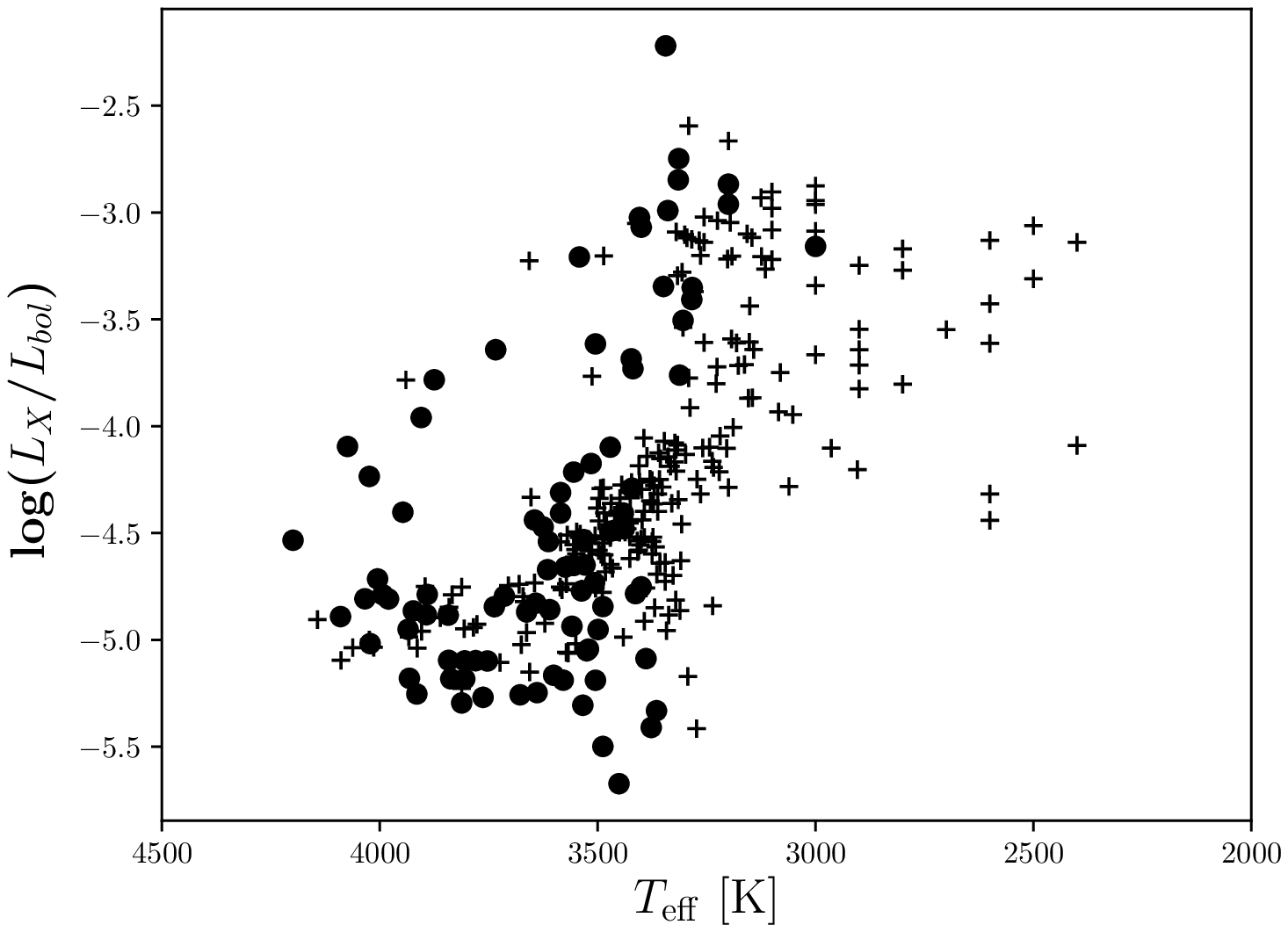}\\
\caption{\label{tefflxlbol} Distribution of $L_{\rm X}/L_{\rm bol}$ as function
  of $T_{\mathrm{eff}}$. Black dots denote $L_{\rm X}/L_{\rm bol}$ measurements
  for our valid sample. Black crosses denote upper limits for $L_{\rm X}/L_{\rm bol}$
for the whole sample, including stars which do not qualify for the valid sample.}
\end{center}
\end{figure}

For the purposes of comparison, we also show in Fig. \ref{tefflxlbol} the
distribution of the $L_{\rm X}/L_{\rm bol}$ values as a function of
the effective temperature $T_{\mathrm{eff}}$ including upper limits
in the $L_{\rm X}/L_{\rm bol}$  measurements.

\section{Spectral series of the \hei\,IR line}\label{appendixa}

\begin{figure*}
\begin{center}
\includegraphics[width=0.5\textwidth, clip]{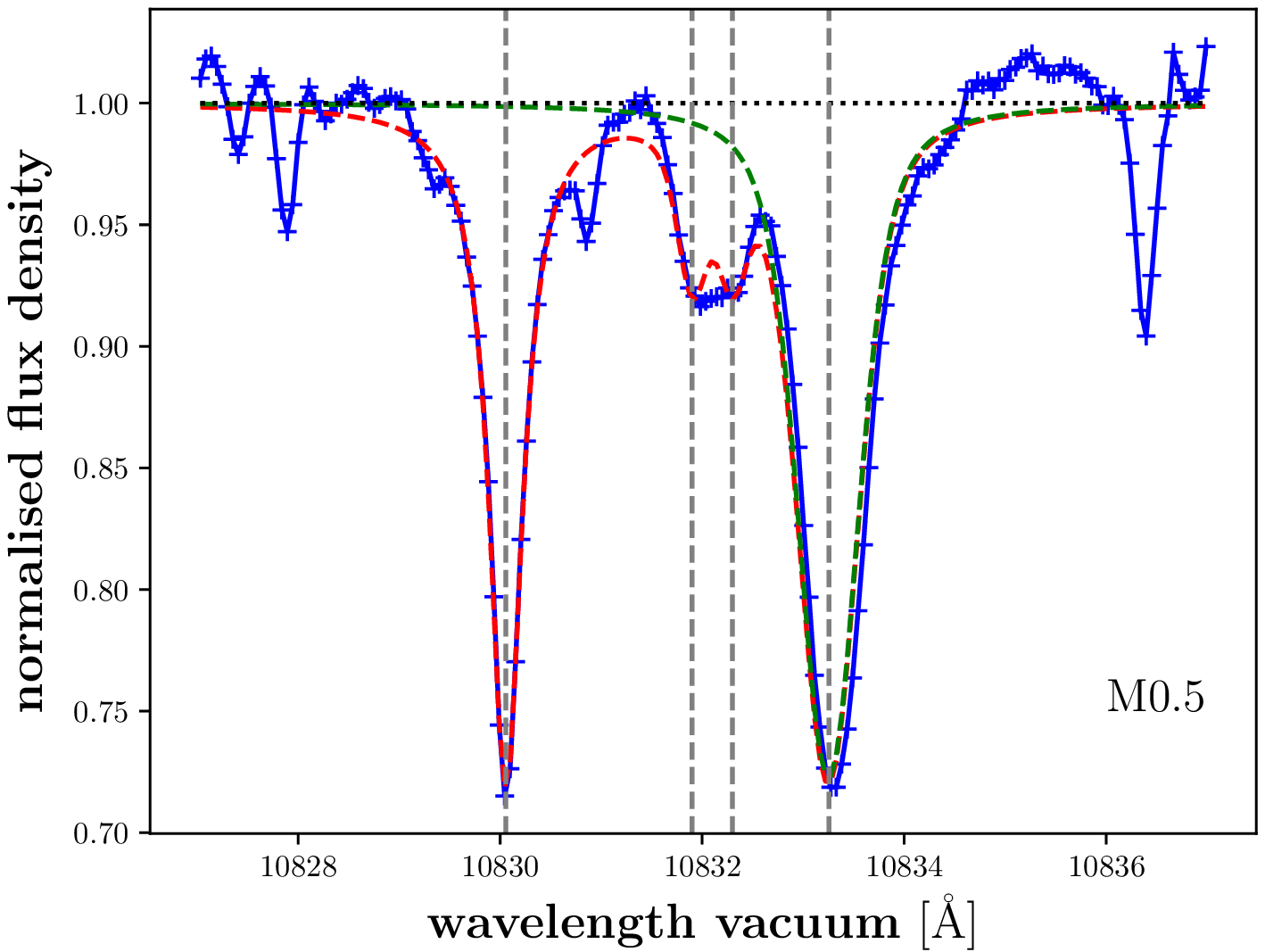}
\includegraphics[width=0.5\textwidth, clip]{gj2servalspec.eps}\\
\includegraphics[width=0.5\textwidth, clip]{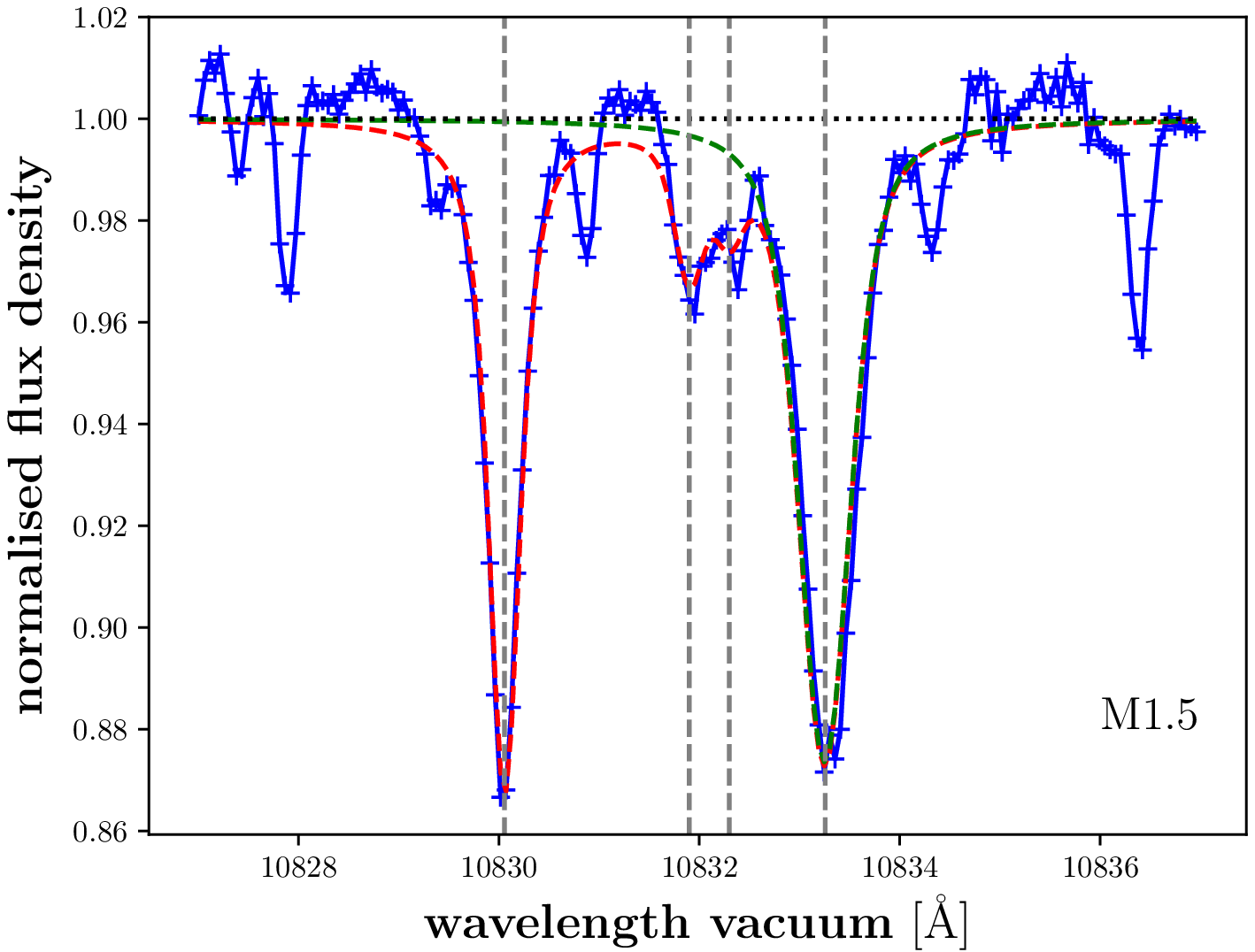}
\includegraphics[width=0.5\textwidth, clip]{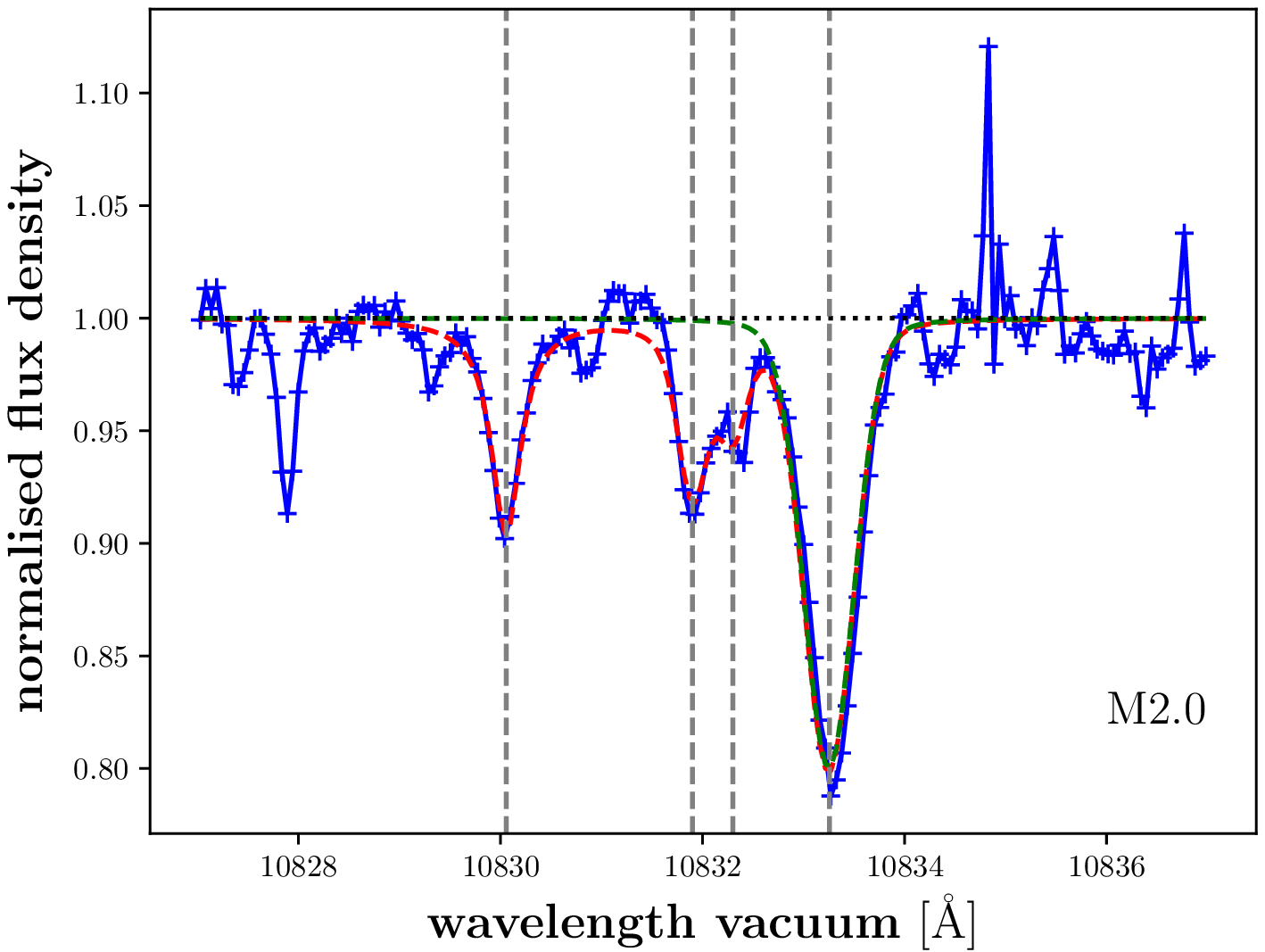}\\
\includegraphics[width=0.5\textwidth, clip]{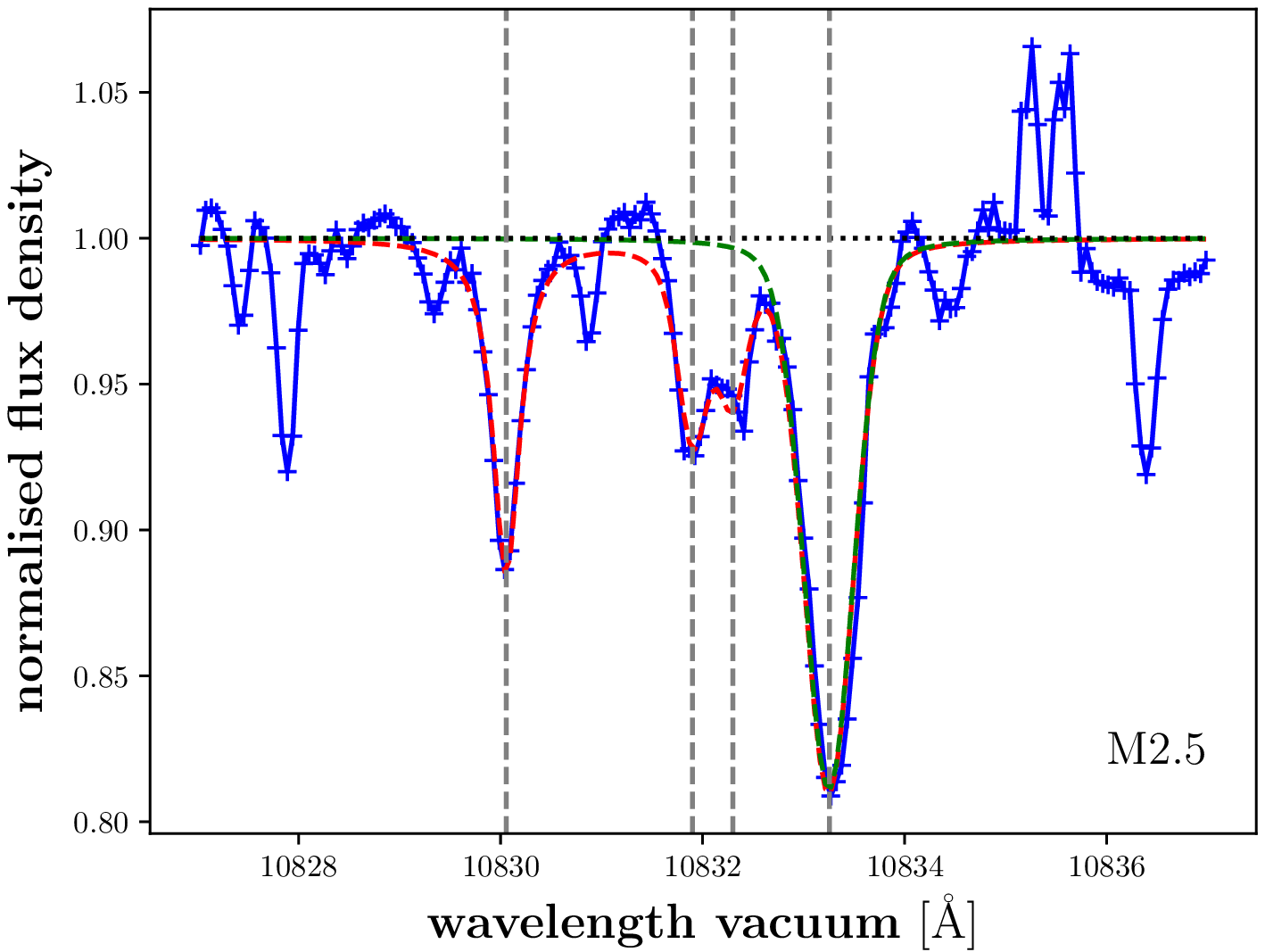}
\includegraphics[width=0.5\textwidth, clip]{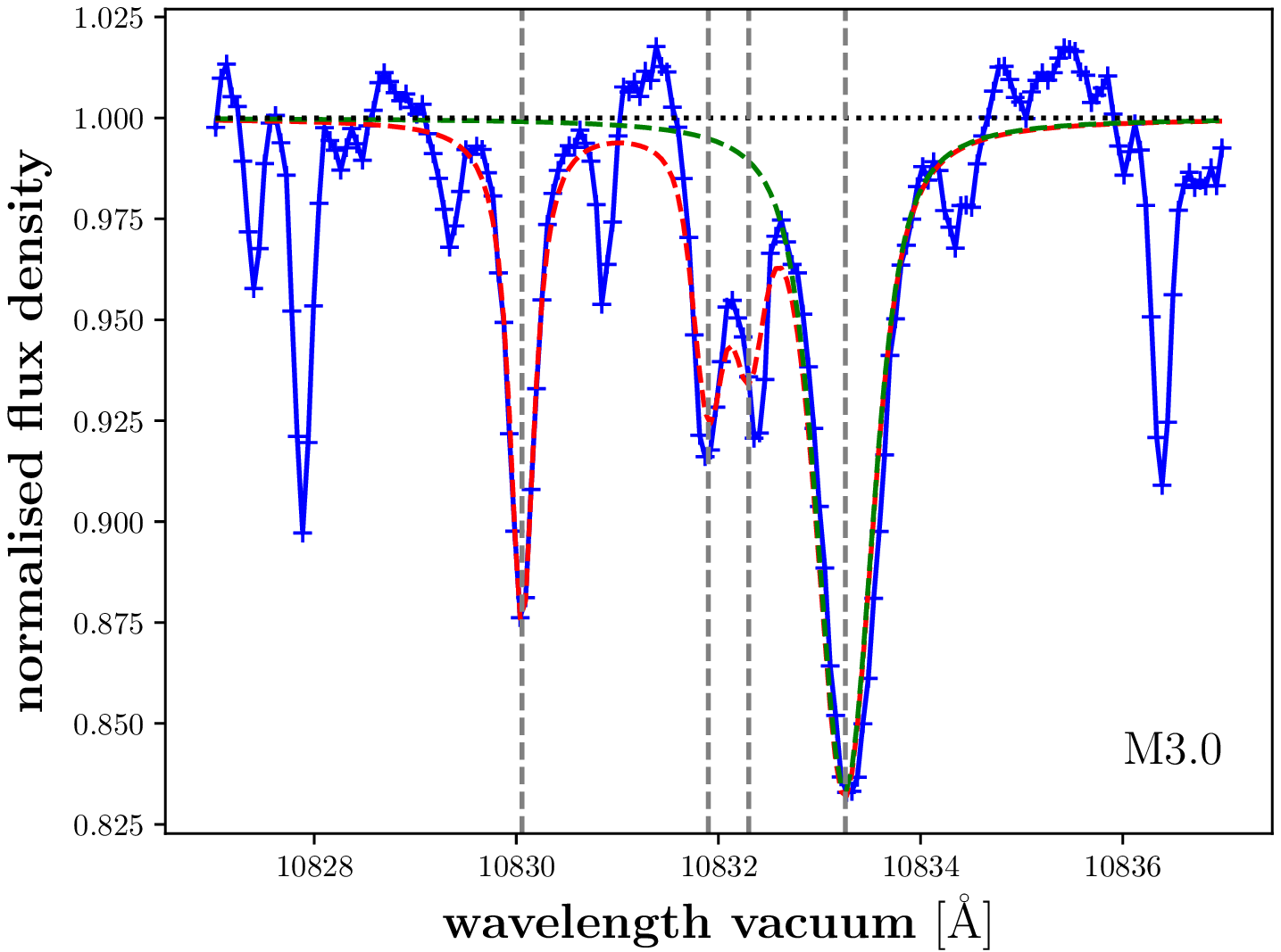}\\
\caption{\label{spectralseries1} Examples 
  for \hei\ IR line and neighbouring spectral region as it changes along with a spectral
  type similar to Fig. \ref{earlyandlateMdwarf}  while also displaying  Voigt fits.
  The observed spectrum is denoted in blue, fit in red, and fit component of the
  \hei\ IR line in green (dashed). Vertical grey dashed lines mark the position of lines
considered in the fit (\ion{Si}{i} line at 10\,830.057 \AA, \hei\,IR line 
at 10\,833.25 \AA, two unidentified lines at 10\,831.9 and 10\,832.3 \AA.)
  \emph{Top left}: M0.5\,V star J02222+478/BD+47~612.
  \emph{Top right}: M1.0\,V star J00051+457/GJ 2.
  \emph{Middle left}: M1.5\,V star J02123+035/BD+02~348.
  \emph{Middle right}: M2.0\,V star J01013+613/GJ~47.
  \emph{Bottom left}: M2.5\,V star J00389+306/Wolf~1056.
  \emph{Bottom right}: M3.0\,V star J02015+637/G~244-047.
}
\end{center}
\end{figure*}

\begin{figure*}
\begin{center}
\includegraphics[width=0.5\textwidth, clip]{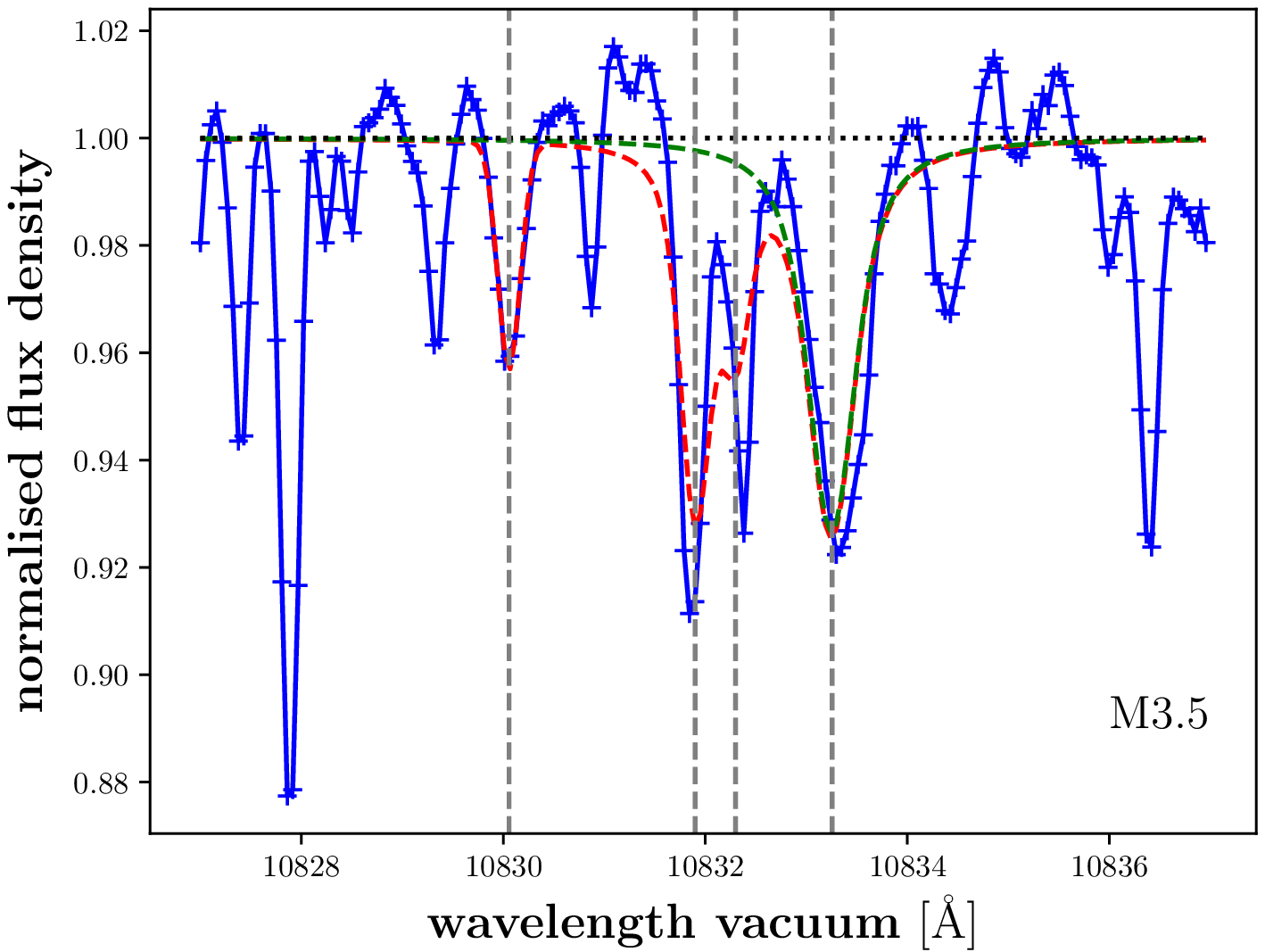}
\includegraphics[width=0.5\textwidth, clip]{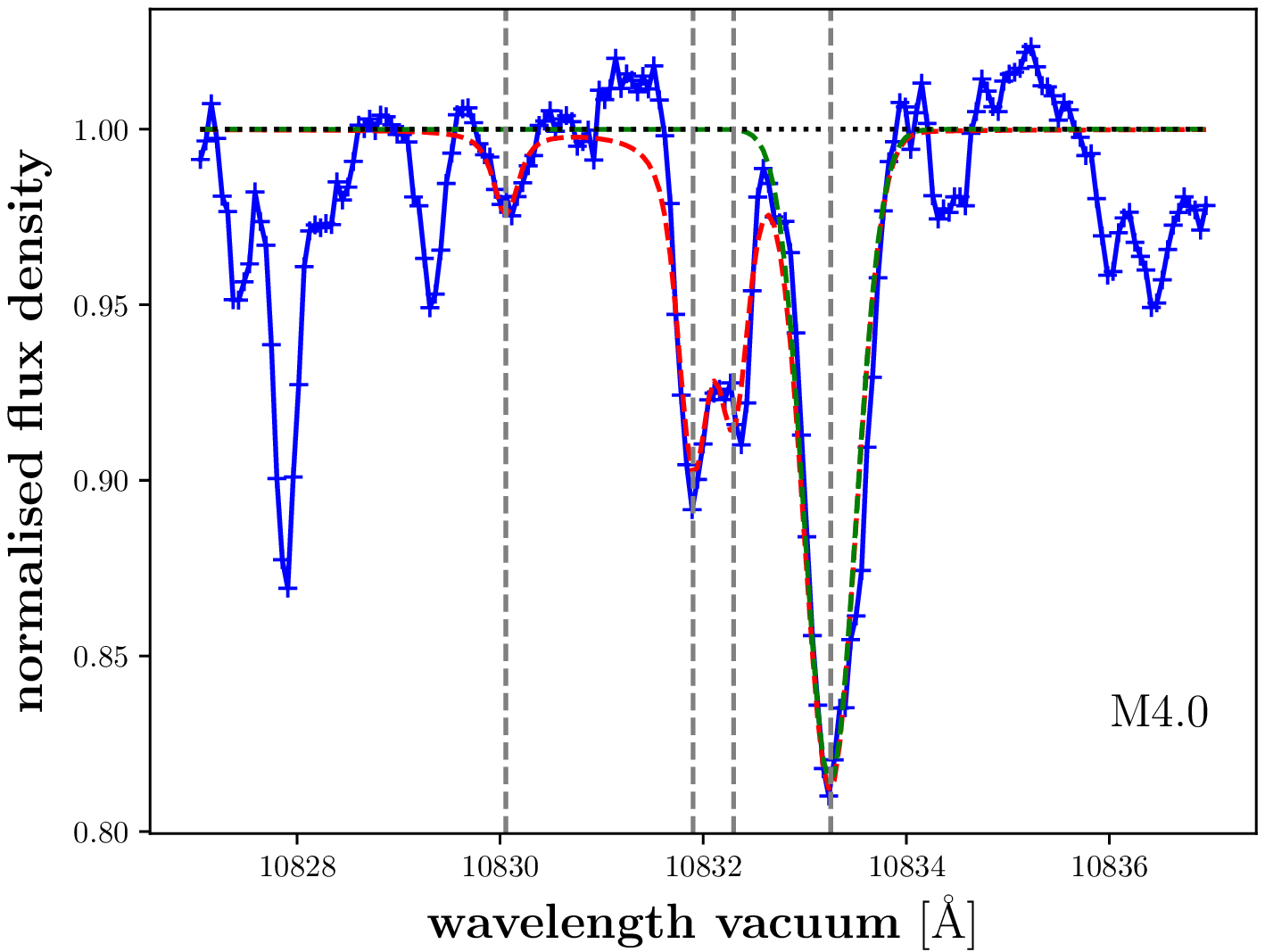}\\
\includegraphics[width=0.5\textwidth, clip]{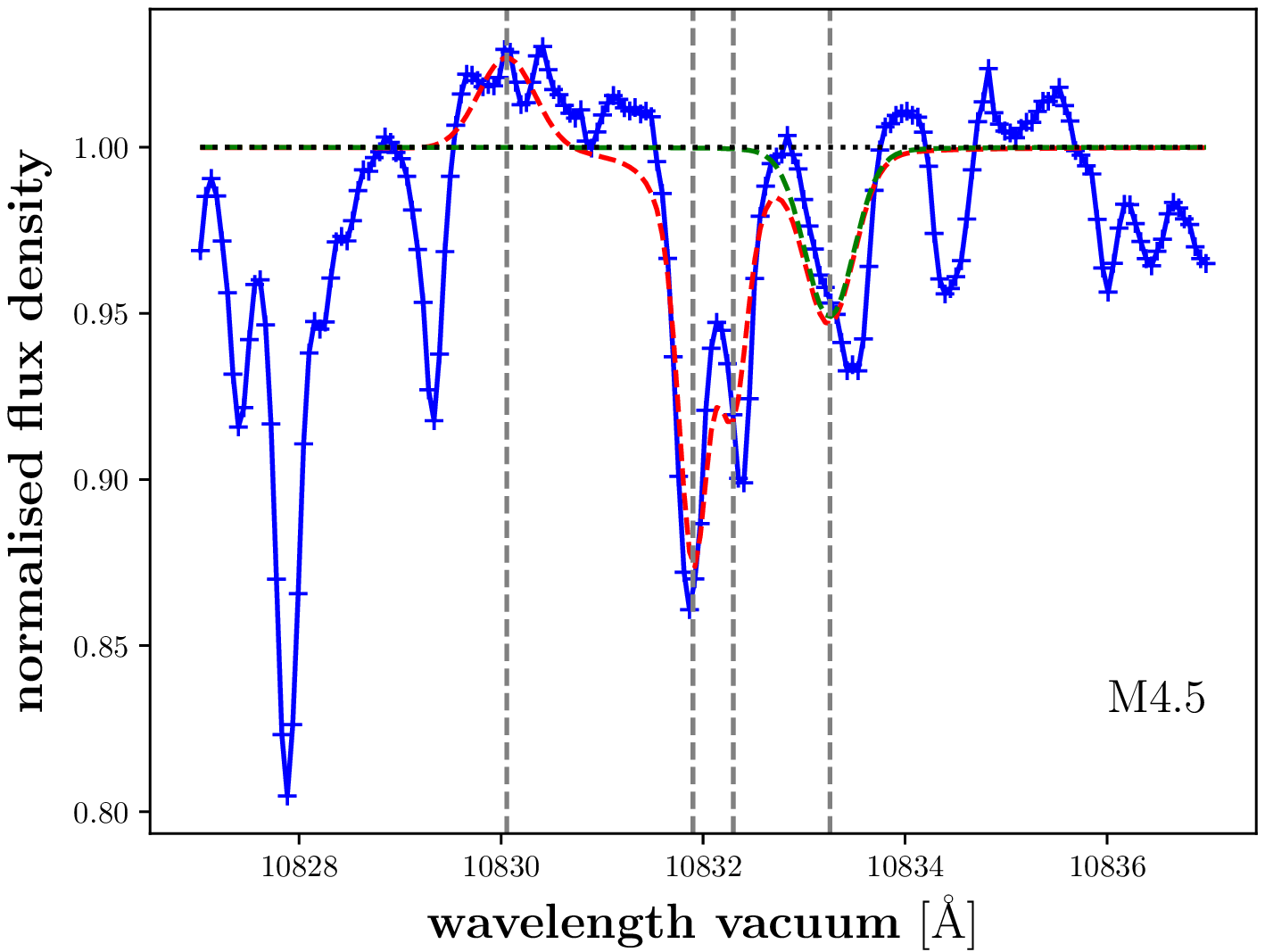}
\includegraphics[width=0.5\textwidth, clip]{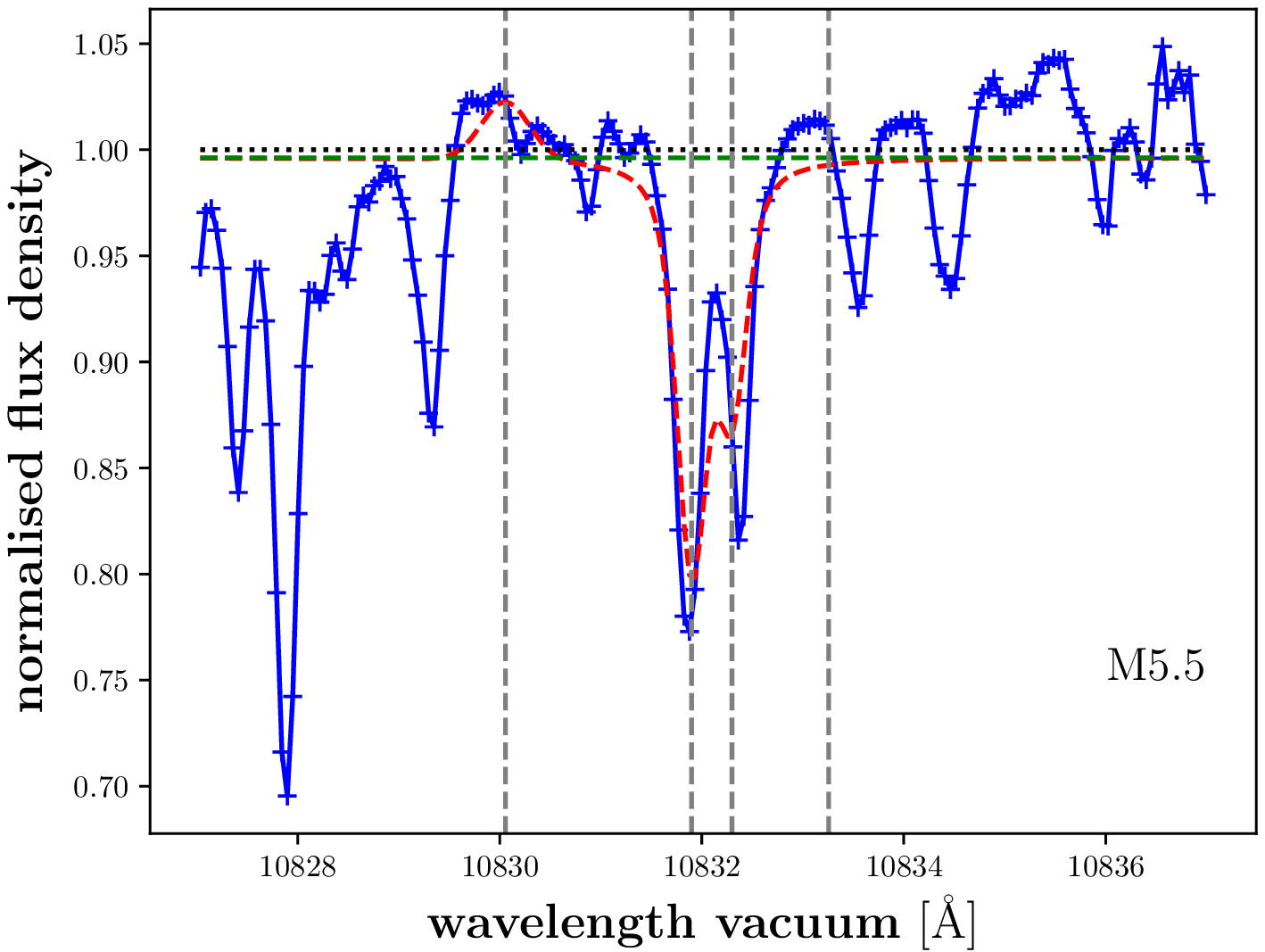}\\
\includegraphics[width=0.5\textwidth, clip]{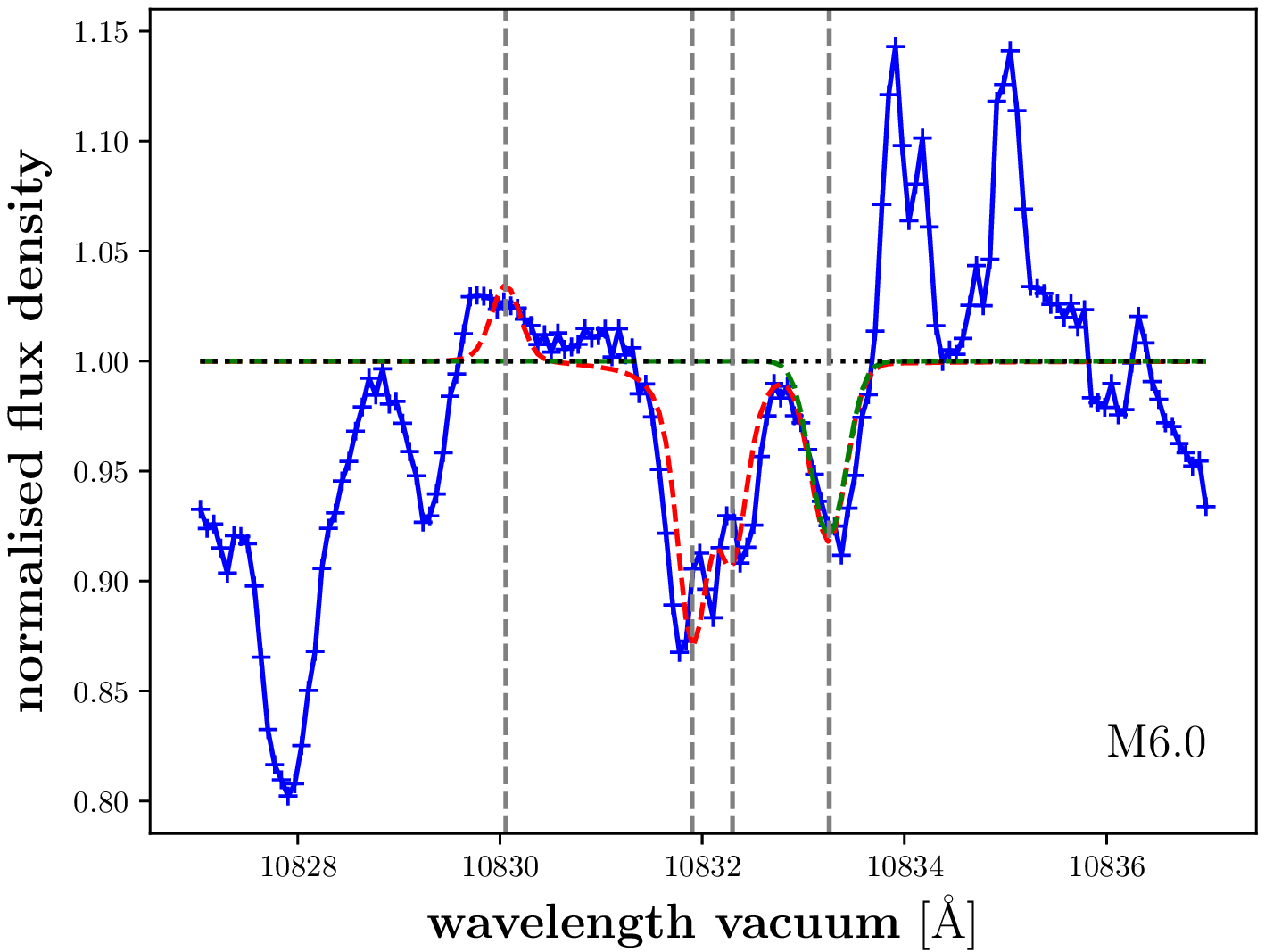}
\includegraphics[width=0.5\textwidth, clip]{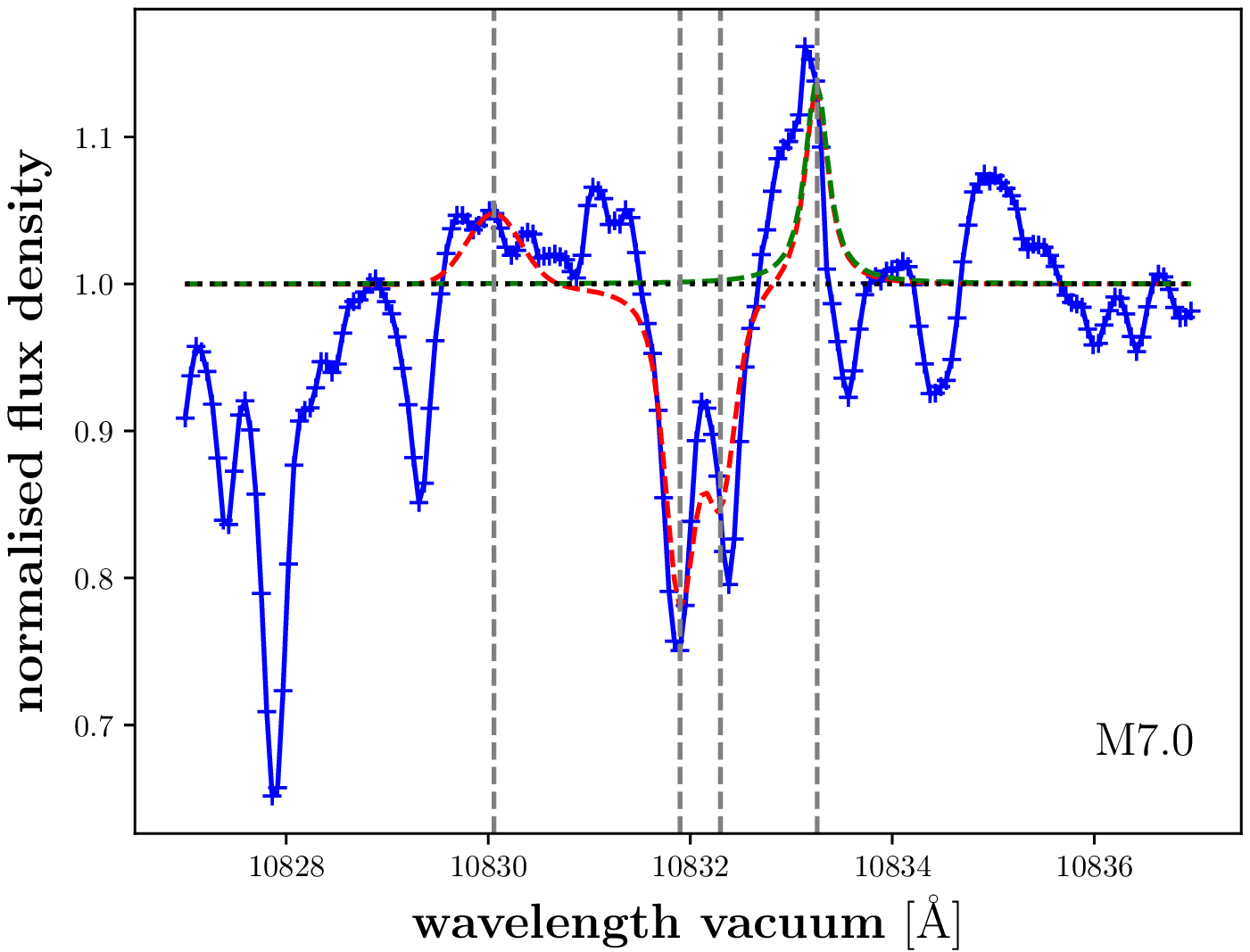}
\caption{\label{spectralseries2} Same as Fig. \ref{spectralseries1}, but for later spectral
  types.
  \emph{Top left}: M3.5\,V star J12479+097/Wolf~437.
  \emph{Top right}: M4.0\,V star J01339-176/LP~768-113.
  \emph{Middle left}: M4.5\,V star J01125-169/YZ Cet.
  \emph{Middle right}: M5.\,V5 star J00067-075/GJ 1002 star.
  \emph{Bottom left}:  M6.0\,V star J14321+081/LP~560-035 showing some emission artefacts
  red-wards of the \hei\, line as well.
  \emph{Bottom right}: M7\,V star J02530+168/Teegarden's star.
}
\end{center}
\end{figure*}

Here we show the development of the \hei\,IR line with spectral type.
In Figs. \ref{spectralseries1} and \ref{spectralseries2} we show the \hei\, IR line
for M0.5\,V to M7.0\,V stars. While the \ion{Si}{i} line has a tendency to diminish towards late
spectral types, the absorption features at 10\,832~\AA\, gain in depth. The \hei\,IR line
also diminishes towards late spectral types and develops some asymmetry with more flux in the red side
from spectral type M3.5\,V on, which apparently belongs to a strengthening feature that is unidentified,
 molecular perhaps,
and also persistent for the very late types where the \hei\,IR line vanishes.

\end{document}